\documentclass[lineno]{JFM-FLM_Au}

\usepackage{subfig}
\usepackage{xcolor}
\usepackage{subcaption}
\usepackage{tcolorbox}
\usepackage{lipsum}
\usepackage{enumitem}  
\usepackage{upgreek}
\usepackage{natbib}
\usepackage{graphicx}
\usepackage[percent]{overpic}
\usepackage{tikz}
\usepackage{float}
\usepackage{booktabs}
\usepackage{caption}
\usepackage{longtable}
\usepackage{tikz}
\usepackage{xcolor}
\usepackage{multirow}
\usepackage{newtxtext}
\usepackage{newtxmath}
\usepackage{hyperref}
\usepackage{pgfplots}
\usepackage{bm}
\usepackage{upgreek}  
\usetikzlibrary{shapes.geometric}
\definecolor{colorstable}{HTML}{FF75BA}
\definecolor{colorunstable}{HTML}{82D3FF}

\usetikzlibrary{calc}

\makeatletter
\@ifundefined{Ta}{}{}
\@ifundefined{Nu}{}{}
\makeatother

\def\x{{\mbox{\boldmath$x$}}}

\def\p{{\mbox{\boldmath$p$}}}

\def\n{{\mbox{\boldmath$n$}}}

\definecolor{blue5}{RGB}{0, 13, 52}
\definecolor{blue4}{RGB}{0, 102, 155}
\definecolor{blue3}{RGB}{0, 137, 204}
\definecolor{blue2}{RGB}{7, 170, 255}
\definecolor{blue1}{RGB}{130, 211, 255}
\definecolor{pink1}{RGB}{255, 117, 186}
\definecolor{pink2}{RGB}{255, 0, 124}
\definecolor{pink3}{RGB}{196, 0, 96}
\definecolor{pink4}{RGB}{153, 0, 76}
\definecolor{pink5}{RGB}{52, 0, 13}
\definecolor{darkgreen}{RGB}{0,130,0}

\def\be{\begin{equation}}
\def\ee{\end{equation}}

\newcommand{\ba}{\begin{eqnarray}}
\newcommand{\ea}{\end{eqnarray}}

\lefttitle{M. Kriening, Z. Yao, M. S. Emran, J. Song, A. Teimurazov and O. Shishkina}
\righttitle{M. Kriening, Z. Yao, M. S. Emran, J. Song, A. Teimurazov and O. Shishkina}

\title{On multiple stable states in Taylor--Couette flow with realistic end-wall boundary conditions}

\author{Marvin Kriening, Zhongzhi Yao, Mohammad S. Emran, Jiaxing Song, Andrei Teimurazov 
\and Olga Shishkina}

\affiliation{Max Planck Institute for Dynamics and Self-Organization, 37077 Göttingen, Germany}

\begin{document}
\renewcommand{\thepage}{\arabic{page}}
\let\oldclearpage\clearpage
\let\oldnewpage\newpage
\let\clearpage\relax
\let\newpage\relax
\vspace*{-1cm}  
\maketitle
\let\clearpage\oldclearpage
\let\newpage\oldnewpage

\begin{abstract}
We investigate Taylor--Couette flow with realistic no-slip boundary conditions at all surfaces 
through direct numerical simulations (DNS) and theoretical analysis. Imposing physically consistent end-wall conditions 
at the top and bottom lids significantly alters the flow dynamics compared to that for periodic boundary conditions. 
We extend the classical angular-momentum-flux framework to account for axial transport, 
which leads to a significantly improved agreement with the Eckhardt--Grossmann--Lohse model 
\citep{eckhardtTorqueScalingTurbulent2007} . \\
A systematic exploration of the parameter space $(\Rey, n) $ uncovers multiple long-lived states with different roll 
number $n$ configurations at identical Reynolds numbers $\Rey$, giving rise to pronounced hysteresis loops occurring under 
realistic boundary conditions. 
Our DNS for no-slip axial end caps reveal a sequence of structural transitions: as the inner-cylinder Reynolds number increases, 
the flow evolves from Taylor vortex flow through chaotic wavy vortex flow and turbulent wavy vortex flow to an axisymmetric 
turbulent Taylor vortex flow. Using modal energy budgets we identify transition mechanisms and quantify how the accessible 
phase-space volume and associated roll-specific angular momentum flux depend on control parameters and the specific flow state.\\
Our findings demonstrate the impact of realistic boundary conditions on the dynamics in Taylor--Couette flow, and how 
they change the stability landscape of multiple states. The coexistence of distinct flow patterns and their stability analysis 
offers promising insights into transition dynamics between laminar and turbulent regimes in closed sheared flows.
\end{abstract}

\newpage

\section{Introduction}
\label{sec:introduction}
Taylor--Couette (TC) flow, the motion of a fluid confined between two independently rotating concentric vertical 
cylinders, has served for over a century as a canonical model for understanding flow instabilities and turbulent dynamics 
in fluid mechanics. 
Its origins trace back to the pioneering experiments of Couette (1890), 
who first used the configuration as a viscometer, and Mallock (1896), who discovered early 
signs of turbulence when rotating the inner cylinder. Taylor's seminal work \citep{taylorStabilityViscousLiquid1923} 
revealed the system's 
linear instability, sparking a long tradition of experimental and theoretical investigation into the transition 
from laminar to turbulent flow. Subsequent studies by \cite{wendtTurbulenteStroemungenZwischen1933} 
and \cite{colesTransitionCircularCouette1965}, and many others have established 
TC flow as a benchmark for exploring fundamental questions in hydrodynamic stability, turbulence, 
and transport processes.

Despite its geometric simplicity, TC flow exhibits a remarkable variety of flow states. 
As the rotation rate of the inner cylinder increases, the base Couette flow gives way to axisymmetric, 
axially periodic, and temporally stationary Taylor vortices. Further increases in the driving parameters trigger 
a sequence of symmetry-breaking transitions. Patterns become increasingly intricate, 
evolving into wavy vortices, modulated structures, and eventually fully developed turbulence. 
Intriguingly, even in the turbulent regime, large-scale coherent structures persist, reflecting an 
interplay between order and disorder \citep{huismanMultipleStatesHighly2014}.

When both cylinders rotate, particularly in the opposite directions, the range of accessible 
flow states expands dramatically, revealing multiple "routes" to turbulence. 
This progressive complexity makes TC flow not only a fertile testing ground for stability theory 
and turbulence modeling but also a bridge between fundamental physics and industrial applications 
\citep{schrimpfTaylorCouetteReactorPrinciples2021, rudelstorferGasLiquidOperations2023}.

Recent advancements in computational fluid dynamics have enabled high-fidelity simulations of TC flow, 
allowing for detailed investigation of turbulent structures and transport phenomena under various boundary conditions. 
Several studies \citep{coughlinTurbulentBurstsCouetteTaylor1996, battenNumericalSimulationsEvolution2002, 
bilsonDirectNumericalSimulation2007, pirroDirectNumericalSimulation2008, brauckmannDirectNumericalSimulations2013, 
ostillaOptimalTaylorCouette2013, ostilla-monicoExploringPhaseDiagram2014} simulated 
turbulent TC flow for relatively high Taylor numbers up to $\mathit{Ta}=4.6\times10^{10}$. Most of these 
simulations \citep{brauckmannDirectNumericalSimulations2013, ostillaOptimalTaylorCouette2013, 
ostilla-monicoExploringPhaseDiagram2014} employed periodic boundary conditions in the axial 
direction. However, imposing no-slip boundary conditions at the horizontal surfaces is crucial for accurately capturing the 
flow dynamics near the walls, where viscous effects are pronounced \citep{jeganathanControllingSecondaryFlows2021}.
Differences between periodic and no-slip end-wall boundary conditions manifest not only in the formation 
of self-organized structures but also in the accessible phase space of temporarily stationary states. 
Realistic boundary conditions enable the system to reach highly asymmetric roll configurations. 
These states not only remain stable but also exhibit significantly higher angular momentum flux compared 
to symmetric states -- a finding that, beyond its fundamental interest for system understanding, highlights substantial 
potential for engineering applications \citep{mamunAsymmetryHopfBifurcation1995,xuDirectNumericalSimulation2023}.

Wall-bounded turbulent flows can exhibit different statistically stationary states even when control parameters remain 
identical \citep{ostilla-monicoExploringPhaseDiagram2014, martinez-ariasEffectNumberVortices2014, 
rameshSuspensionTaylorCouette2019}. However, this phenomenon is 
not unique to Taylor--Couette flow -- it has also been observed in other convective systems such as Rayleigh--Bénard 
convection \citep{ahlersHeatTransferLarge2009, vanderpoelConnectingFlowStructures2011, wangMultipleStatesTurbulent2020}, 
tilted convection \citep{wangMultipleStatesHeat2018}, double diffusive convection \citep{yangMultipleStatesTransport2020} 
and magnetoconvection \citep{McCormack_Teimurazov_Shishkina_Linkmann_2025} to name a few. Despite being known for several 
decades, this topic requires further investigation, as we are still far from a complete understanding despite its 
profound significance for numerous fields where wall-bounded turbulence plays a central 
role.

In this study, we perform direct numerical simulations (DNS) of realistic TC flow for no-slip boundary conditions at all 
surfaces, with the outer cylinder fixed and the rotation rate of the inner cylinder systematically varied.
A $\Rey$ range of $90\leq \Rey\leq 7500$ is covered by mainly focusing on system geometries, which are 
characterized by $\Gamma = 11$ and $\Gamma=30,$ respectively.
The latter geometry has been chosen as our main investigation setup due to the broad spectrum for comparison 
with experimental data by \cite{martinez-ariasEffectNumberVortices2014}, which covers a wide $\Rey$ range 
characterized by several different flow states.
We hence study multiple stable states in this geometry by systematically varying the inner Reynolds number $\Rey_i.$

\begin{figure}[h!]
    \centering
\begin{tikzpicture}

\node[anchor=south west,inner sep=0] (image) at (0,0) {
\includegraphics[width=0.9\textwidth]{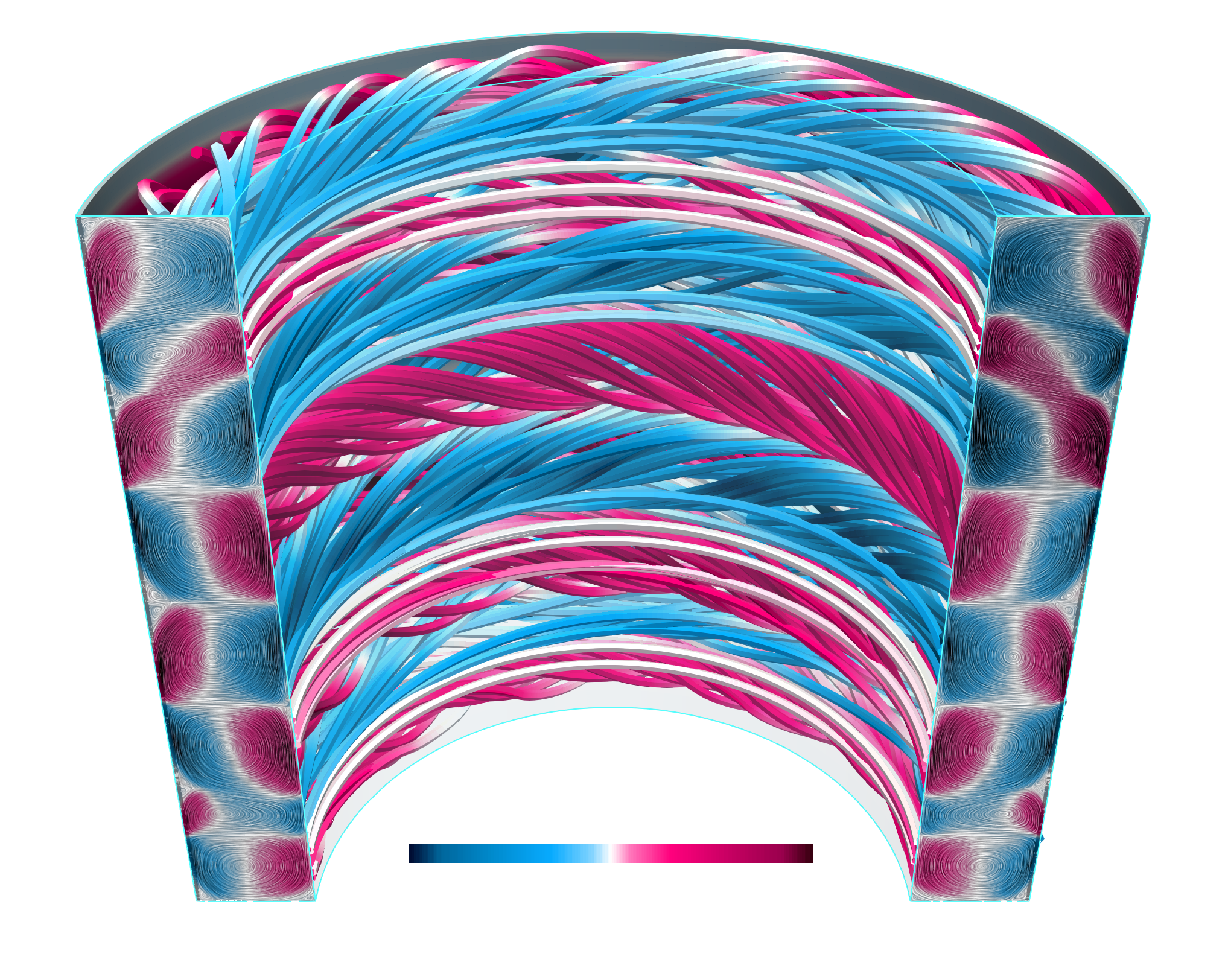}
        };

\begin{scope}[x={(image.south east)},y={(image.north west)}]
\draw[<->, thick, black, line width=1pt] 
                (0.74,0.057) -- (0.84,0.057) node[midway,below] {$d$};
\draw[<->, thick, black, line width=1pt] 
                (0.8825,0.089) -- (0.98,0.77) node[midway,right] {$H$};
\node[fill=white, inner sep=4pt, minimum width=3.6cm, minimum height=0.3cm] at (0.5,0.18) {};
\node at (0.37,0.095) {\small $-0.18$};
\node at (0.64,0.095) {\small $0.18$};
\node[yshift=-1.15cm] at (0.5,0.1875) {$\displaystyle u_z/(r_o\omega_i)$};

\draw[->, thick, black, line width=1.0pt] 
    (0.48,0.175) + (0.04,0) arc (-45:225:0.03);
\node at (0.5,0.19) {$\omega_i$};
\end{scope}
\end{tikzpicture}
\vspace{-0.25cm}
\caption{Example of a TC flow as obtained in the DNS for $\Rey = 5600$, $\Gamma = 2\pi$, $\eta = 5/7$ as in the 
    setup of \cite{ostillaOptimalTaylorCouette2013}, but with solid top and bottom lids, 
    illustrated by isosurfaces of the axial velocity $u_z.$ Only half of the computational domain is shown.}
\label{fig:geometry_3d_ostilla}
\end{figure}
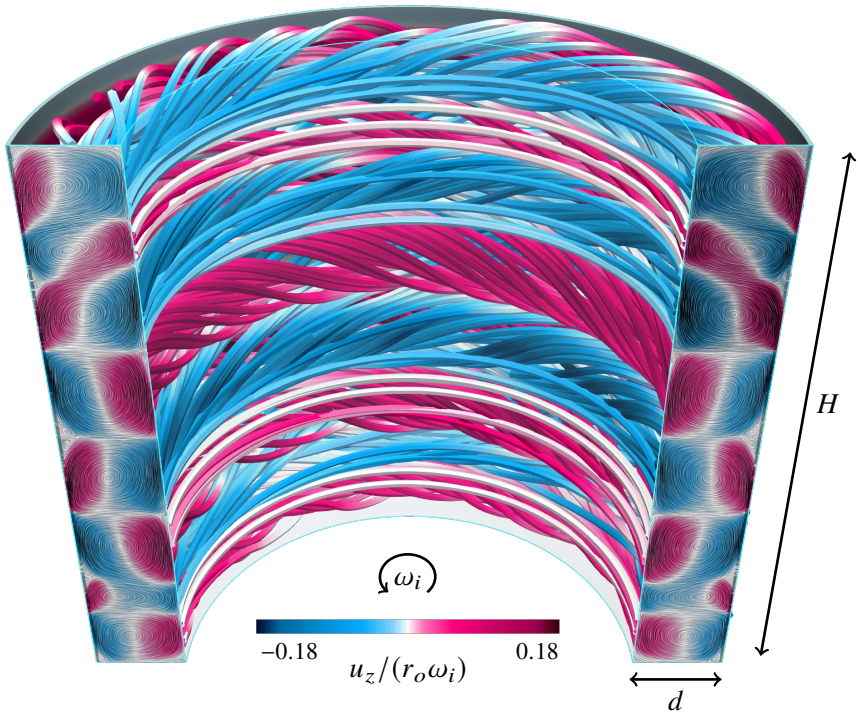

The paper is organized as follows. Section \ref{sec:system} describes the Taylor--Couette system, governing 
equations, control and response parameters. 
Section \ref{sec:numerics} details our numerical implementation, including the treatment of 
wall-lid singularities and validates our numerical approach against experimental data. Sections \ref{sec:DNS_analysis} 
and \ref{sec:multiple_states} present the DNS results.
Section \ref{sec:DNS_analysis} highlights changes in the global and local dynamics due to the realistic boundary conditions, while
section \ref{sec:multiple_states} presents multiple stable states and hysteresis phenomena and
discusses the underlying physical mechanisms. Section \ref{sec:conclusion} 
concludes with a summary of our findings and future challenges.

\newpage 

\section{System Description}
\label{sec:system}
\subsection{Control parameters and governing equations}

We consider the flow of an incompressible Newtonian fluid with kinematic viscosity $\nu$ confined 
between two concentric vertical cylinders of radii $r_i$ (inner) and $r_o$ (outer) with height $H$. 
The inner cylinder rotates with angular frequency $\omega_i$ while the outer cylinder remains 
stationary ($\omega_o = 0$). 
Upper and lower plates rotate with the outer sidewall at the same angular frequency
and therefore remain at rest within our system of consideration. 
The system is characterized by three dimensionless parameters: the radius ratio $\eta$, the aspect ratio $\Gamma$, and the 
Reynolds number $\Rey$,

\begin{equation}
    \eta = \frac{r_i}{r_o}, \quad \Gamma = \frac{H}{d}, \quad \Rey \equiv \Rey_i = \frac{\omega_i r_i d}{\nu},
\end{equation}

\noindent where $d = r_o - r_i$ is the gap width. Another useful quantity we use is the 
Taylor number $\mathit{Ta}$ \citep{eckhardtTorqueScalingTurbulent2007}
\begin{equation}
    \mathit{Ta} \equiv \left(\frac{r_o+r_i}{2}\right)^6\frac{d^2}{r_o^2 r_i^2 \nu^2} (\omega_o-\omega_i)^2.
\end{equation}
\noindent The Navier--Stokes equations for incompressible TC flow in a rotating frame read:
\begin{equation}
\nabla\cdot\mathbf{u}=0,\,\,\frac{\partial \mathbf{u}}{\partial t} + (\mathbf{u} \cdot \nabla) \mathbf{u} = -\frac{1}{\rho} \nabla p + \nu \nabla^2 \mathbf{u} -2\boldsymbol{\Omega}\times\mathbf{u}, \label{eq:TC_leading_eq}
\end{equation}

\noindent where $\mathbf{u}$ is the velocity field, $p$ is the pressure, $\rho$ is the density, $\nu$ is the 
kinematic viscosity, $\boldsymbol{\Omega}=\omega_o \mathbf{e_z}$ with $\omega_o$ the reference angular frequency, 
and $\mathbf{e_z}$ the unit vector in the axial direction.

An example of an analyzed setup is shown in figure \ref{fig:geometry_3d_ostilla}, which presents a cross-section of our 
Taylor--Couette system along with the axial flow field obtained from our DNS.

\noindent In cylindrical coordinates $(r,\phi,z)$, and under the assumption of a non-moving outer cylinder, they take the 
following form:
\begin{align}
    \partial_t u_r +(\mathbf{u}\cdot\nabla)u_r - \frac{u_\phi^2}{r} &= -\frac{\partial_r p}{\rho} + \nu\left(\nabla^2 u_r - \frac{u_r}{r^2} -\frac{2}{r^2}\partial_\phi u_\phi\right) \label{eq:ns_r}, \\
    \partial_t u_\phi +(\mathbf{u}\cdot\nabla)u_\phi + \frac{u_\phi u_r}{r} &= -\frac{1}{r\rho}\partial_\phi p + \nu\left(\nabla^2 u_\phi - \frac{u_\phi}{r^2} +\frac{2}{r^2}\partial_\phi u_r\right) \label{eq:ns_phi}, \\
    \partial_t u_z +(\mathbf{u}\cdot\nabla)u_z &= - \frac{\partial_z p}{\rho} + \nu\nabla^2 u_z \label{eq:ns_z},\\
    0 &= \frac{1}{r}\partial_r(r u_r)+\frac{1}{r}\partial_\phi u_\phi +\partial_z u_z \label{eq:continuity},
\end{align}

\noindent where $\mathbf{u} \equiv (u_r, u_\phi, u_z).$ We impose no-slip conditions at all boundaries:
\begin{alignat}{4}
    &\text{Inner cylinder: } \hspace{0.5cm}& u_r &= 0,\hspace{0.25cm} & u_\phi &= \uppsi r_i\omega_i,\hspace{0.25cm} & u_z &= 0 \quad \text{at } r = r_i \notag \\
    &\text{Outer cylinder: } \hspace{0.5cm}& u_r &= 0,\hspace{0.25cm} & u_\phi &= 0,\hspace{0.25cm} & u_z &= 0 \quad \text{at } r = r_o \notag \\
    &\text{Top/bottom lids: } \hspace{0.5cm} & u_r &= 0,\hspace{0.25cm} & u_\phi &= 0,\hspace{0.25cm} & u_z &= 0 \quad \text{at } z = 0, H \label{eq:bc_lids},
\end{alignat}
\noindent where a smoothing function $\uppsi$ (see section \ref{sec:numerics}) is applied to 
avoid singularities at $r=r_i$ and $z=0$ or $z=H$. These boundary conditions differ from the usually 
used periodic approximation in the sense that they introduce an explicit $z$-dependence even in the 
mean flow due to constraints on the azimuthal velocity $u_\phi$ which must 
satisfy $u_\phi(r,\phi,0) = u_\phi(r,\phi,H) = 0$ at the lids.

\subsection{Response characteristics and boundary-layer thicknesses}

The seminal work of Eckhardt, Grossmann, and Lohse \citep{eckhardtTorqueScalingTurbulent2007} established the foundation for 
understanding transport dynamics in TC flow. By averaging the azimuthal velocity $u_{\phi}$ 
over a cylindrical surface of height $H$ and area $A(r) = 2\pi rH$ coaxial with the rotating cylinders, 
where $r_i\leq r \leq r_o$, they derived equations describing angular momentum transport and kinetic energy 
dissipation within TC flow under the assumption of periodic boundary conditions at the top and bottom walls. However, 
for no-slip boundary conditions at the top and bottom lids, an extended approach is required. 
The azimuthal velocity $u_\phi$ becomes a function not only of radius $r$ but also of axial position $z$ due 
to the influence of Ekman layers \citep{czarny} and the confinement imposed by the walls limiting vertical extent.

To derive the modified angular-momentum-flux, we consider equations
(\ref{eq:ns_phi}) and (\ref{eq:continuity}) under time and area averaging over a cylindrical surface at fixed radius.
Following \cite{eckhardtTorqueScalingTurbulent2007}, we introduce radial and axial fluxes of angular velocity: 
\begin{align}
    J_{r}(r,z) &= r^3\left(\langle u_r\omega\rangle_{\phi,t}-\nu\partial_r\langle\omega\rangle_{\phi,t} \right), \notag \\
    J_{z}(r,z) &= r^3\left(\langle u_z\omega\rangle_{\phi,t}-\nu\langle\partial_z(\omega)\rangle_{\phi,t}\right),
\end{align}

\noindent where $\omega\equiv u_\phi/r$ denotes the angular frequency and $\langle .\rangle_{\phi,t}$ is an average 
in time as well as in azimuthal direction.
Under the assumption that the flow is axisymmetric and statistically stationary, we obtain the following mean azimuthal 
angular momentum equation:
\begin{equation}
    \partial_r J_r(r,z) + \partial_z J_z(r,z) = 0 \label{eq:main_relation},
\end{equation}
\noindent which means, that the flux vector $(J_r, J_z)$ is divergence-free in the $(r,z)$-plane. 
The average of (\ref{eq:main_relation}) in the vertical $z$-direction yields: 
\begin{equation}
    \partial_r\langle J_r(r,z)\rangle_{z} + (J_z(r,H) - J_z(r,0))/H = 0 \label{eq:main_2},
\end{equation} 
where
\begin{equation}
    J_z(r,H) = -\nu r^3\partial_z\langle\omega\rangle_{\phi,t}\Big|_{z=H}, \quad J_z(r,0) = -\nu r^3\partial_z\langle\omega\rangle_{\phi,t}\Big|_{z=0}.
\end{equation}
In the case of periodic axial boundary conditions one obtains $J_z(r,H) = J_z(r,0),$ 
so that $\langle J_r(r,z)\rangle_z=\text{const.}$ in $r$ \citep{eckhardtTorqueScalingTurbulent2007}.
Here, $\langle .\rangle_{z}$ denotes an average in the $z$-direction. However, for no-slip end-walls, 
Ekman boundary layers at $z=0$ and $z=H$ generate axial transport of angular momentum through the top and bottom 
lids, and $\langle J_r(r,z)\rangle_z$ is not radially constant
anymore. Instead, using equation (\ref{eq:main_2}), we can obtain the radially constant quantity, 
which is the radial flux connected by the 
cumulative axial exchange with the top and bottom plates, 
\begin{align}
    J^\omega(r) &\equiv \langle J_r(r,z)\rangle_{z} + \frac{1}{H}\int_{r_i}^r\left(J_z(r',H) - J_z(r',0)\right)\mathrm{d}r' \notag \\
    &= \langle J_r(r,z)\rangle_{z} - \frac{\nu}{H}\int_{r_i}^r r'^3\left[\partial_z\langle\omega\rangle_{\phi,t}\Big|_{z=H} - \partial_z\langle\omega\rangle_{\phi,t}\Big|_{z=0}\right]\mathrm{d}r' \notag \\
    &= \text{const. in } r. \label{eq:J_omega}
\end{align}
\noindent We also call $J^\omega(r)$ the total angular momentum flux. It consists of two contributions:\\
the classical radial transport term $J_r(r,z)$ that appears also in periodic systems, and an 
additional axial transport term arising from the $z$-dependence of the azimuthal velocity 
induced by the no-slip endwalls. In order to ensure meaningful validation with experiments, 
the total angular momentum flux $J^\omega(r)$ will further be normalized by its value in the 
conductive flow regime for periodic boundary conditions \citep{eckhardtTorqueScalingTurbulent2007}
\begin{equation}
    J^\omega_0(r)=2\nu r_i^2 r_o^2 \frac{\omega_i - \omega_o}{r_o^2-r_i^2}.
\end{equation}
Since $J^\omega$ is independent of radius (conserved quantity), and $J^\omega(r_i) = J^\omega(r_o)$ we can 
relate the velocity gradients at the inner and outer cylinders as follows:

\begin{equation}
    \frac{\mathrm{d}\langle\omega\rangle_{r,z,t}}{\mathrm{d}r}\Bigg|_{r_o} = \eta^3\frac{\mathrm{d}\langle\omega\rangle_{r,z,t}}{\mathrm{d}r}\Bigg|_{r_i} - \frac{1}{r_o^3H}\int_{r_i}^{r_o}r'^3\left(\partial_z\langle\omega\rangle_{\phi,t}\Big|_{z=H} - \partial_z\langle\omega\rangle_{\phi,t}\Big|_{z=0}\right)\mathrm{d}r'. \label{eq_bc_thickness}
\end{equation}

\noindent We further introduce boundary layer approximations
\begin{equation}
    \frac{\mathrm{d}\langle\omega\rangle_{r,z,t}}{\mathrm{d}r}\Bigg|_{r_i}\approx -\frac{\Delta_i}{\delta_i}, \quad \frac{\mathrm{d}\langle\omega\rangle_{r,z,t}}{\mathrm{d}r}\Bigg|_{r_o}\approx -\frac{\Delta_o}{\delta_o}, \label{eq:bc_thick}
\end{equation}
\noindent where $\Delta_i=|\omega_i-\overline{\omega}|$ and $\Delta_o=|\overline{\omega}|$ are the angular 
velocity differences between, respectively, the inner ($\omega_i$) and outer ($\omega_o$) walls and 
the bulk flow ($\overline{\omega}$), and $\delta_i (\delta_o)$ is the viscous boundary layer thickness 
at the inner (outer) cylinder. From (\ref{eq_bc_thickness}) and (\ref{eq:bc_thick}) we obtain an approximate formulation
for the ratio between the outer and inner boundary layer thicknesses,
\begin{equation}
    \frac{\delta_o}{\delta_i} = |\overline{\omega}| \left(\eta^3\Delta_i + \delta_i r_o^{-3}H^{-1}\int_{r_i}^{r_o}r'^3\left(\partial_z\langle\omega\rangle_{\phi,t}\Big|_{z=H} - \partial_z\langle\omega\rangle_{\phi,t}\Big|_{z=0}\right)\mathrm{d}r'\right)^{-1}. \label{eq:bl_thickness}
\end{equation}

\noindent The term in brackets represents the axial correction to the classical periodic 
prediction \citep{eckhardtTorqueScalingTurbulent2007}, accounting for the 
$z$-dependent flow structure near the endwalls.

\section{Numerical setup}
\label{sec:numerics}
Direct numerical simulations were conducted with the in-house code \textsc{goldfish} 
\citep{shishkinaThermalBoundaryLayer2015,reiterGenerationZonalFlows2021, reiterFlowStatesHeat2022}, which 
has been adapted to account for rotation \citep{hornToroidalPoloidalEnergy2015, hornProgradeRetrogradeOscillatory2017, 
zhangBoundaryZonalFlow2020, zhangBoundaryZonalFlows2021}, internal wall inside the convection container 
\citep{shishkinaThermalBoundaryLayer2015, emranNaturalConvectionCylindrical2020},
and different boundary conditions \citep{eckeConnectingWallModes2022, zhangWallModesTransition2024}, and has been 
widely used in previous studies of convective flows.
\textsc{goldfish} employs a fourth-order finite-volume discretization on staggered grids 
and a third-order Runge--Kutta time-marching scheme.

With the reference length $r_o,$ reference time $1/\omega_i,$ and reference velocity $r_o\omega_i$ the non-dimensionalized 
governing equations (\ref{eq:TC_leading_eq}), under assumption of pure inner cylinder rotation, take the form 
\begin{equation*}
\nabla' \cdot \mathbf{u}' = 0,
\end{equation*}
\begin{equation}
\frac{\partial \mathbf{u}'}{\partial t'} + (\mathbf{u}' \cdot \nabla') \mathbf{u}' = -\nabla' p' + \frac{\kappa}{\sqrt{T\!a}} \nabla'^2 \mathbf{u}',
\end{equation}
\noindent where
\begin{equation*}
\kappa = \frac{(1+\eta)^3(1-\eta)}{8\eta}.
\end{equation*}
\noindent In the following we omit the prime symbols for simplicity. 

A significant numerical challenge arises at the junction between the rotating inner cylinder and the 
stationary top and bottom lids, where the no-slip condition creates a velocity discontinuity. 
Without appropriate treatment, this singularity causes numerical instabilities due to extreme velocity gradients.
We address this issue by introducing a smooth transition function in the axial direction. Near the lids 
at $z = 0$ and $z = H$, we modify the azimuthal velocity boundary condition on the inner cylinder:
\begin{equation}
    u_\phi(r_i, z) = r_i\omega_i\uppsi(z, \epsilon),
\end{equation}

\noindent where $\uppsi$ is a smoothing function and $\epsilon$ a constant that controls the transition width. 
In this regard, a sigmoid function has been employed (see section \ref{seq:appendix}), which minimizes non-physical 
second derivatives, that affect viscous terms, while 
ensuring smooth first derivatives. 

To ensure reliable numerical results, both temporal and spatial resolution must be sufficiently small to 
resolve Kolmogorov length and time microscales \citep{popeTurbulentFlows2000}.
Our first resolution requirement is $\updelta / \eta_K < 1.0$, 
where $\updelta = \sqrt[3]{r \updelta_\phi \updelta_r \updelta_z}$ is the local grid spacing and $\eta_K$ is the Kolmogorov 
length. Additionally, following \cite{ostillaOptimalTaylorCouette2013}, we check the flatness of  
the radial profile of the angular momentum flux.
Since angular momentum flux should remain constant across the gap width (as derived in section \ref{sec:system}), we 
require the peak-to-peak variation to be below $1\%$.
This criterion, combined with the Kolmogorov-scale resolution, yields the grid configurations used throughout this work.
In this study, we were able to achieve sufficient angular momentum flux constancy for all cases presented while also 
maintaining close adherence to the Kolmogorov criterion (see Appendix \ref{seq:appendix_resolution}).

\section{DNS results, experimental comparison and boundary layer estimations}
\label{sec:DNS_analysis}
\subsection{Effect of smoothing range}
\begin{figure}[h!]
    \centering
    \includegraphics{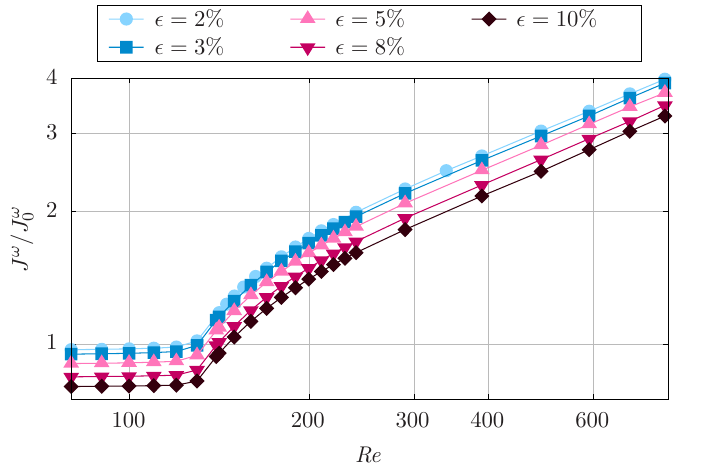}
    \caption{Angular momentum flux development for different smoothing ranges 
    ($\epsilon=2\%$, 3\%, 5\%, 8\% and 10\%) in the range of $50 \leq \Rey \leq 700$ for $\Gamma=30$ and $\eta=0.909$.}
    \label{fig:nusselt_smoothing}
\end{figure}

Systematic variation of the smoothing width $\epsilon$ reveals the following observations. 
The angular momentum flux exhibits a constant offset across all Reynolds numbers, 
representing a proportional shift in $J^\omega$, while the fundamental flow physics remain unchanged.
Figure \ref{fig:nusselt_smoothing} demonstrates this behavior across the parameter 
range $\Rey \in [50, 700]$ for five different smoothing widths ($\epsilon = 2\%, 3\%, 5\%, 8\% \text{ and } 10\%$ of 
the domain height). A plausible explanation for the constant offset lies in the effective thickening of the 
Ekman boundary layers near the top and bottom lids. The smoothing zones damp velocity 
fluctuations in these regions, leading to increased viscous dissipation and reduced 
turbulent momentum transport. This effect mimics a thicker viscous sublayer, which is 
known to reduce the efficiency of angular momentum transport in wall-bounded flows 
\citep{poncetTurbulentCouetteTaylor2013, lengAspectratioDependenceHeat2022}. Since the additional dissipation scales 
proportionally with the local momentum flux, the relative reduction remains approximately constant across 
different Reynolds numbers.

We furthermore observe a shift in the onset of Taylor vortex flow in comparison to the 
experiments chosen for validation. 
The shift in critical Reynolds number arises because the smoothing function creates 
artificial rotation of the lids near the junction with the cylinder walls, effectively 
modifying the boundary conditions \citep{wendtTurbulenteStroemungenZwischen1933}.

These systematic offsets necessitate appropriate corrections 
in both Reynolds number and angular momentum flux. 
We calibrate this effect using the $\epsilon = 2\%$ case as our reference configuration, which has been chosen to 
replicate typical experimental conditions \citep{vangilsTorqueScalingTurbulent2011}, while maintaining computationally 
feasible mesh requirements.
The observed offsets are purely 
additive in Reynolds number ($\Rey \to \Rey + \Delta\Rey_{\text{offset}}$) and purely 
multiplicative in angular momentum flux ($J^\omega \to J^\omega (1 + \varepsilon_{\text{offset}})$), 
where $\Delta\Rey_{\text{offset}}$ was chosen to be exactly $10\%$ of the Reynolds number value, 
where convection sets in. This value lies within the bifurcation shift predicted by \citet{wendtTurbulenteStroemungenZwischen1933} 
for the aspect ratios examined in the present study, leading to quantitative agreement with experimental data.
$\varepsilon_{\text{offset}}$ has been selected such that our angular momentum flux calculations match the experimental 
results for Reynolds numbers in the conductive regime of \cite{ostillaOptimalTaylorCouette2013, 
rameshSuspensionTaylorCouette2019} or \cite{martinez-ariasEffectNumberVortices2014}.

\noindent This procedure enables straightforward systematic corrections that allow consistent comparison with experimental data. 
Importantly, since these offsets do not affect scaling 
exponents or the relative locations of regime transitions when properly normalized, the smoothing approach 
does not alter the fundamental flow physics but merely shifts the absolute values of control and response parameters.

\subsection{The role of initial conditions}
The choice of initial conditions is crucial in complex hydrodynamic systems exhibiting multiple 
stable states \citep{ruelleNatureTurbulence1971, fenstermacherDynamicalInstabilitiesTransition1979, 
wangMultipleStatesTurbulent2020, yaoDirectNumericalSimulations2025}. We have observed significant differences in flow 
dynamics depending on the initialization procedure, as evidenced by validation of our DNS with the measurements by 
\cite{martinez-ariasEffectNumberVortices2014} (see Section \ref{sec:validation}).

\begin{figure}[h!]
    \centering
    \begin{tikzpicture}
        \node[anchor=south west,inner sep=0] (image) at (0,0) 
            {\includegraphics[scale=0.25, trim=5cm 0 10cm 0, clip]{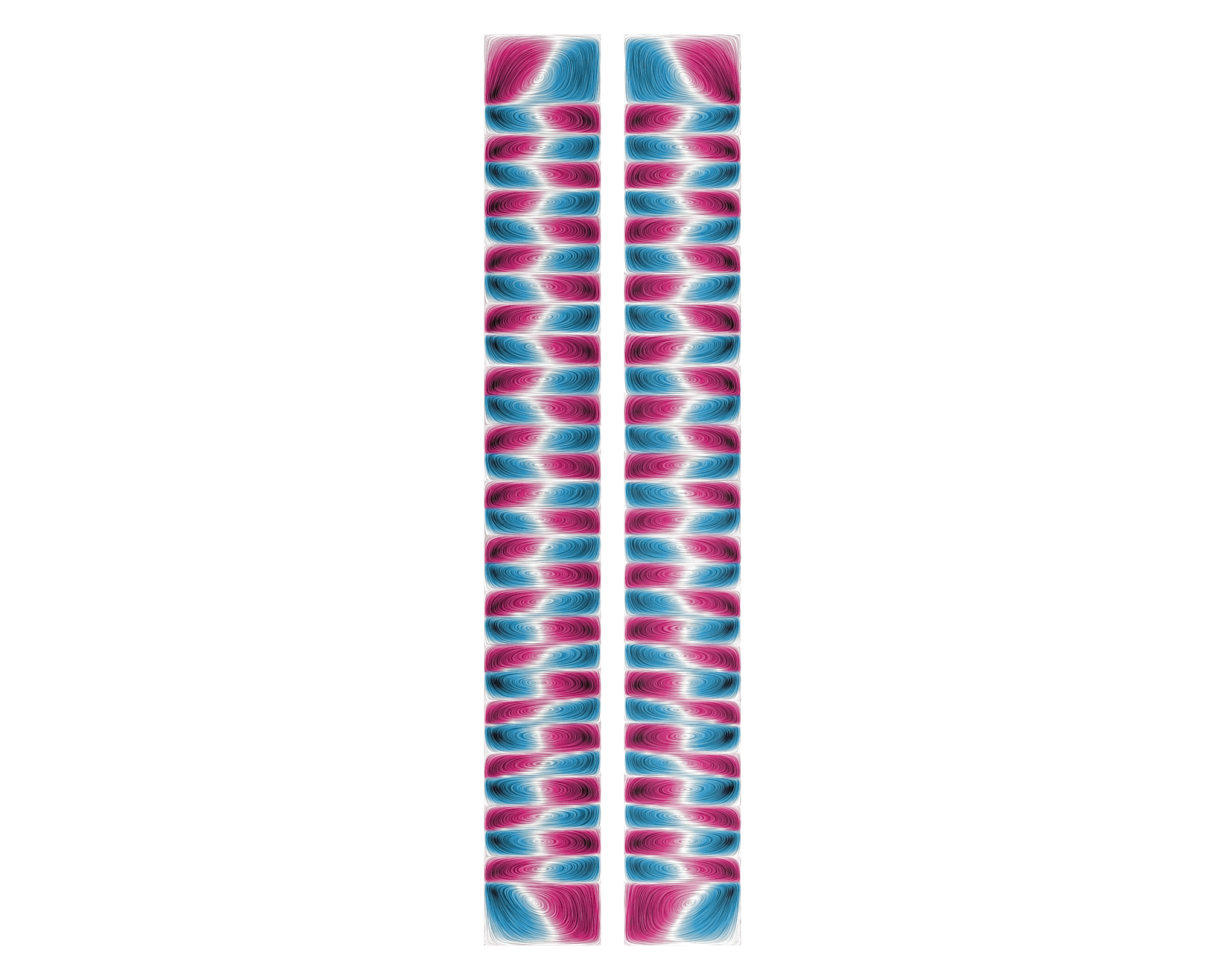}};
        \node[anchor=north west, yshift=5pt, xshift=1.7cm] at (image.north west) {(\textit{a})};
        
        \begin{scope}[x={(image.south east)},y={(image.north west)}]
            \draw[-stealth, thick] (0.30,0.05) -- (0.42,0.05) node[below, scale=0.9] {$r$};
            \draw[-stealth, thick] (0.30,0.05) -- (0.30,0.17) node[left, scale=0.9] {$z$};
        \end{scope}
    \end{tikzpicture}
    \hspace{-0.1\textwidth}
    \begin{tikzpicture}
        \node[anchor=south west,inner sep=0] (image) at (0,0) 
            {\includegraphics[scale=0.25, trim=5cm 0 10cm 0, clip]{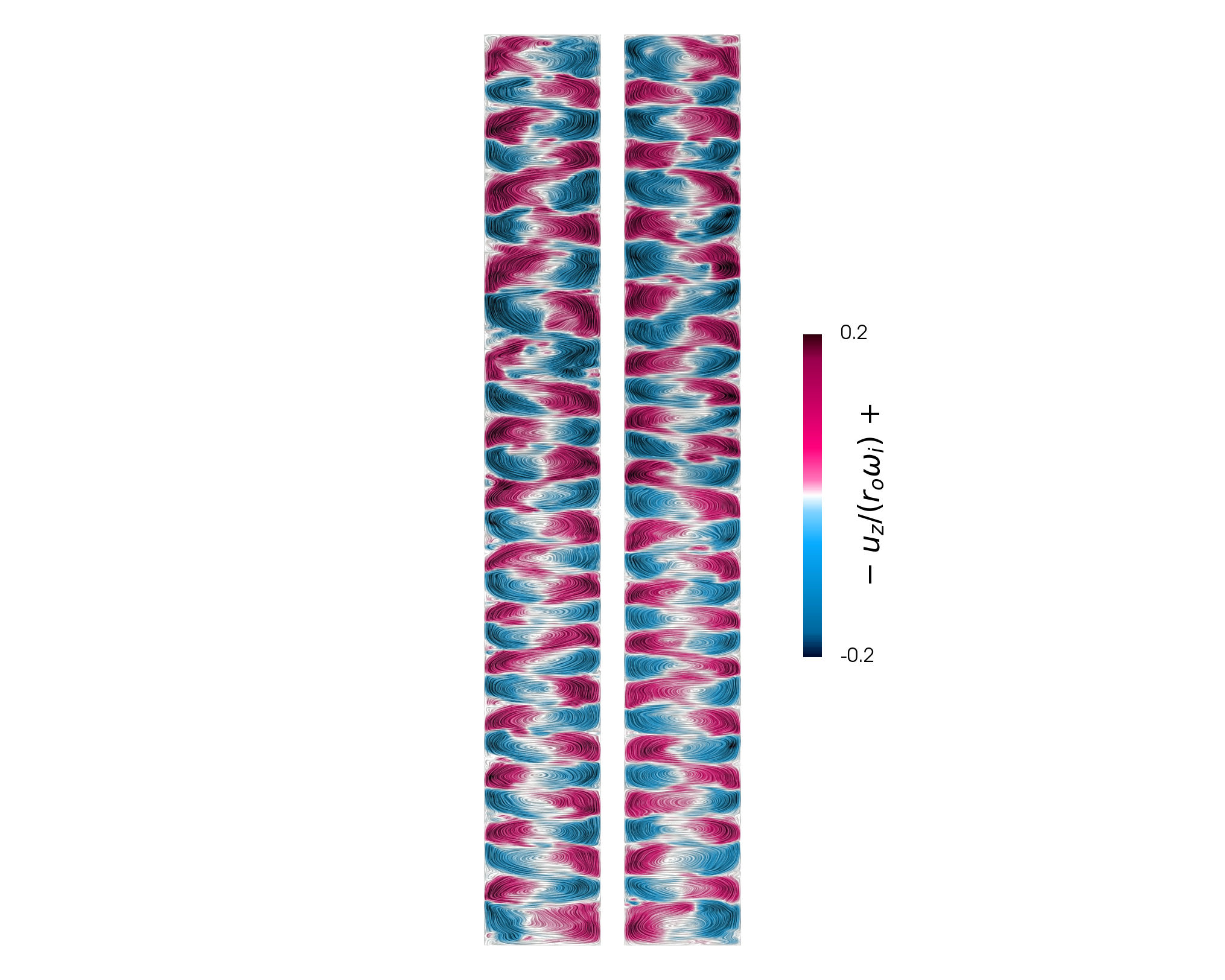}};
        \node[anchor=north west, yshift=5pt, xshift=1.7cm] at (image.north west) {(\textit{b})};
        
        \begin{scope}[x={(image.south east)},y={(image.north west)}]

            \node[fill=white, inner sep=2pt, minimum width=1.2cm, minimum height=5cm] at (1.03,0.6) {};
            
            \node at (0.99,0.675) {$\phantom{-}0.2$};
            
            \node[rotate=90] at (1.0,0.5) {$\displaystyle u_z/(r_o\omega_i)$};
            
            \node at (0.99,0.325) {$-0.2$};

        \end{scope}
    \end{tikzpicture}
    \caption{Instantaneous snapshots of axial velocity within the Taylor--Couette system in central vertical 
    cross-sections for $\Gamma = 30, \eta = 0.909$ and different $\Rey$ (\textit{a}) $\Rey = 1500$ 
    and (\textit{b}) $\Rey = 4500$, initialized with $\mathbf{u}_{\text{initial}}=\mathbf{0}$. 
    Presented is a TVF state on the left and a WVF state on the right.}
    \label{fig:flowfields_2d}
\end{figure}
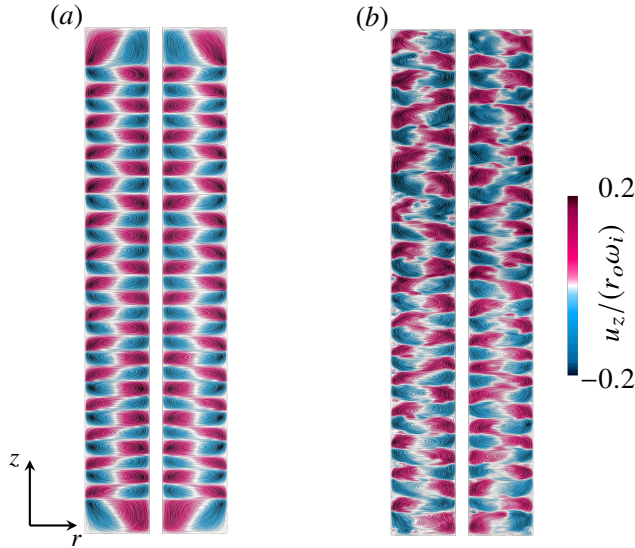

Simulations initialized with zero velocity fields consistently evolve toward conditionally 
stable Taylor vortex flow states with asymmetric roll configurations (see figure \ref{fig:flowfields_2d}), 
even at high inner cylinder rotation rates where alternative flow states might be expected. 
However, by adding white noise perturbations (see section \ref{seq:appendix}) to these simulation results, 
we observed significant dynamical shifts. 
Specifically, systems underwent transitions from Taylor vortex flow (TVF) to wavy vortex flow (WVF) in regimes 
where WVF has been observed experimentally and predicted theoretically \citep{martinez-ariasEffectNumberVortices2014}.

These observations motivated our initialization strategy for a systematic study of multiple states in TC flow. 
In our DNS, we initialize the flow with an analytically constructed velocity field that satisfies the continuity 
equation and boundary conditions while containing the characteristic Taylor vortex structure 
(see section \ref{seq:appendix}). This perturbed vortex flow state for a specific roll configuration serves as the initial condition 
for our systematic exploration of the multiple states in phase space (for details see section \ref{sec:multiple_states}). 
\\\noindent The excellent agreement of our angular momentum flux results with experimental 
data (see section \ref{sec:validation}) validates this initialization procedure.

\subsection{Comparison with experiments}
\label{sec:validation}
\begin{figure}[h!]
    \centering
    \hspace{-1cm}
    \includegraphics{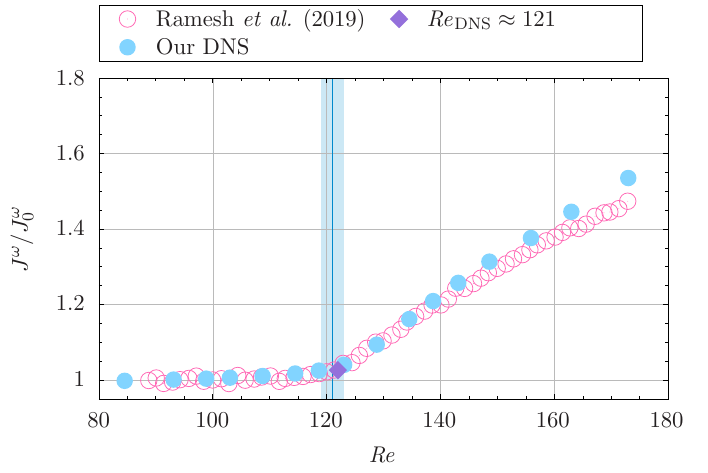}
    \vspace{-0.2cm}
    \caption{Angular momentum flux development for moderate $\Rey$ values in the setup of 
    \cite{rameshSuspensionTaylorCouette2019} for $\Gamma=11$, $\eta=0.914$. The blue region highlights the onset of 
    convection in the experiments, 
    which agrees perfectly with our simulation result (\textcolor[HTML]{9370DB}{\Large$\blacklozenge$}).
    The earlier transition in comparison to the periodic 
    assumption is due to perturbations by Ekman vortices (see figure \ref{fig:flowfields_ramesh}).}
    \label{fig:ramesh_comparison}
\end{figure}

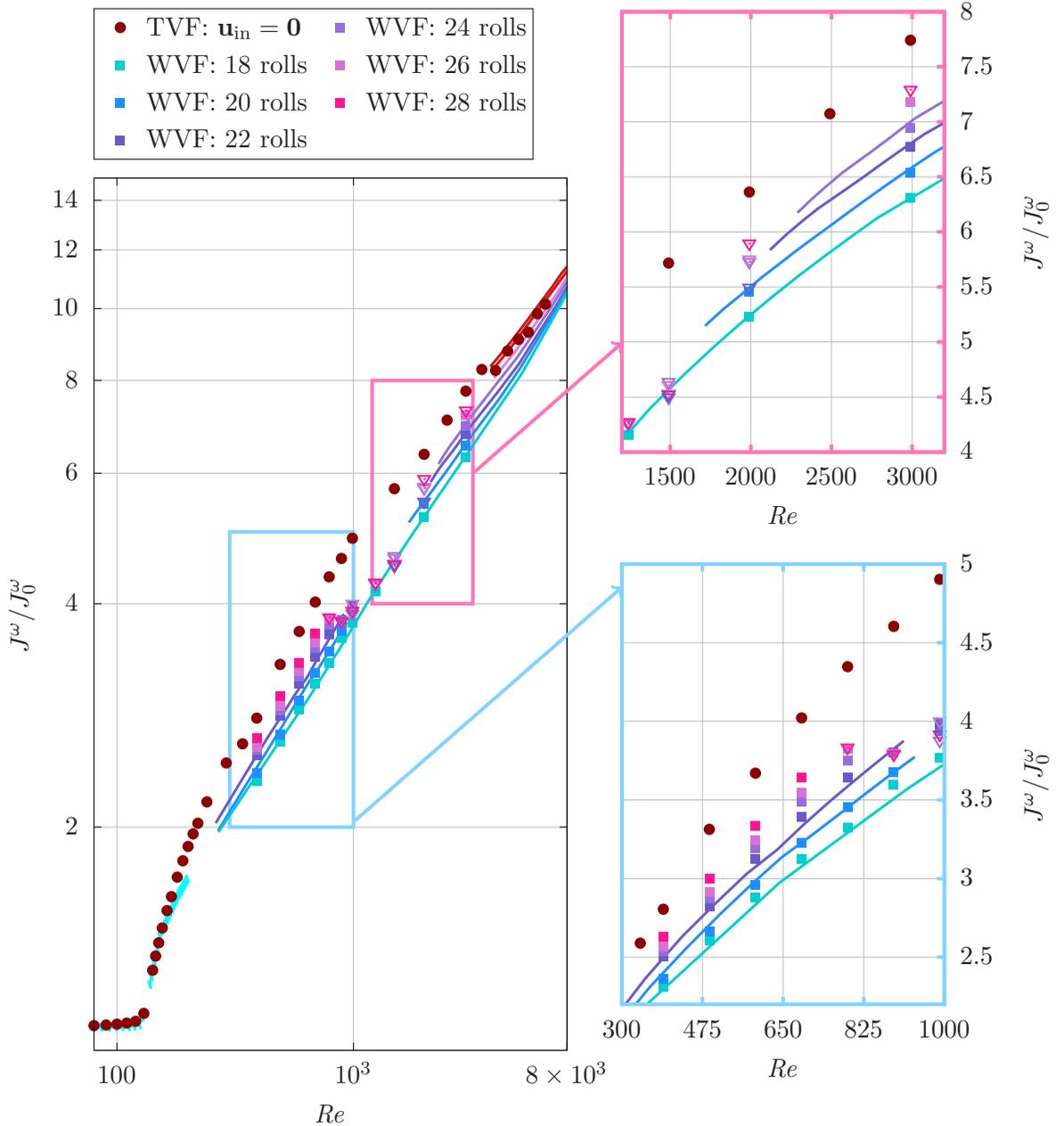
\begin{figure}[h!]
    \centering
    \hspace*{-3cm}
    \begin{minipage}[c]{3.74in}
        \centering
        \tikz[remember picture] \node[inner sep=0pt] (mainplot) {\includegraphics{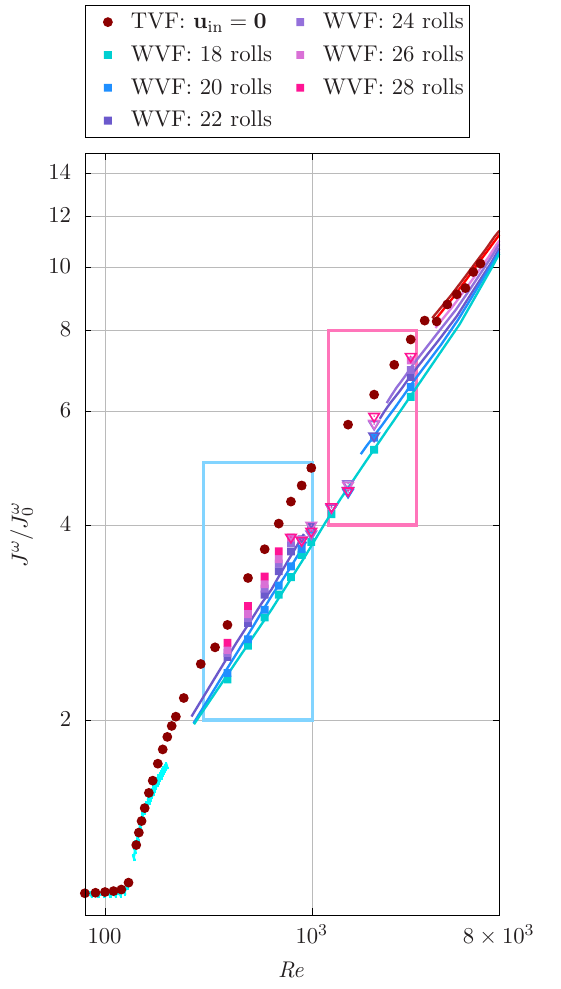}};
    \end{minipage}
    \hspace*{-1cm}
    \begin{minipage}[c]{1.45in}
        \centering
        \vspace{-0.1in}
        \tikz[remember picture] \node[inner sep=0pt] (inset2) {\includegraphics{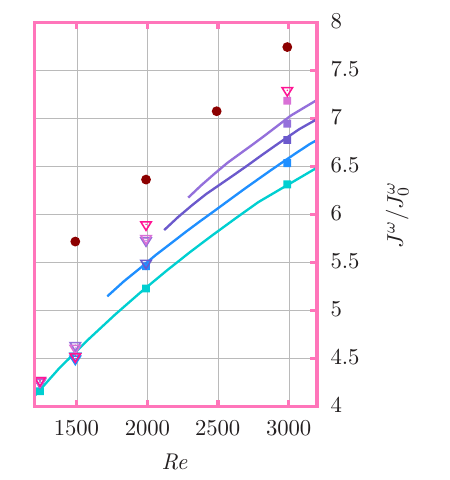}};
        \vspace{0.26in}
        \tikz[remember picture] \node[inner sep=0pt] (inset1) {\includegraphics{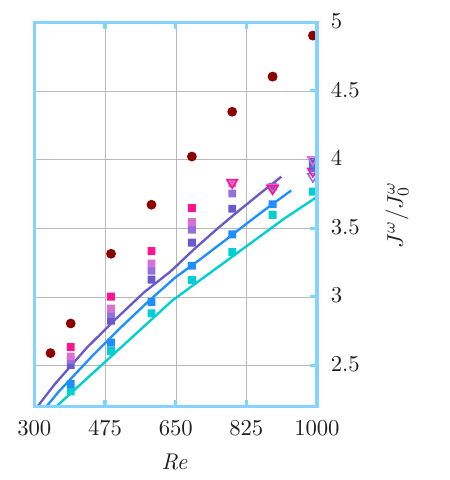}};
    \end{minipage}

    \begin{tikzpicture}[remember picture, overlay]

    \draw[->, line width=1.5pt, colorstable] 
        ($(mainplot.east) + (-2.45, 1.4)$) -- ($(inset2.west) + (0.6, -1.2)$);
    
    \draw[->, line width=1.5pt, colorunstable] 
        let \p1 = ($(mainplot.east) + (-2.45, 1.4)$),
            \p2 = ($(inset2.west) + (0.6, -1.2)$),
            \p3 = ($(mainplot.east) + (-4.22, -3.75)$),
            \p4 = ($(inset1.west) + (0.6, 0)$),
            \n1 = {\x2-\x1},  
            \n2 = {\y2-\y1},  
            \n3 = {\x4-\x3},  
            \n4 = {\n3*\n2/\n1}  
        in
        ($(mainplot.east) + (-4.22, -3.75)$) -- 
        ($(\p3) + (\n3, \n4)$);
    \end{tikzpicture}
    
    \vspace{-0.2cm}
    \caption{Angular momentum flux development for $50 \leq \Rey \leq 8 \times 10^3$ in the 
    setup with $\Gamma=30$ and $\eta=0.909$. 
    Solid lines denote the experimental data by \cite{martinez-ariasEffectNumberVortices2014} and symbols our DNS data for 
    different stable roll configurations. Brown circles show our DNS results for zero initial velocity, 
    whereas coloured squares (triangles) show our DNS results 
    of multiple persistent (non-persistent) state simulations with an initial velocity specifically 
    chosen in regard of the respective roll 
    configuration (see section \ref{seq:appendix}). We can highlight an excellent agreement with the 
    experimental results for the multiple states 
    phase space region, with a reduction of possible persistent configurations 
    between $800 < \Rey < 2 \times 10^3$ and a sudden regrowth of the accessible phase
    space volume for higher $\Rey$. This transition can be understood by analysing the 
    transition in the flow dynamics with rising $\Rey$. Closer views are shown on the right figures.}
    \label{fig:martinez}
\end{figure}

We compare our numerical results with two experimental studies: \cite{rameshSuspensionTaylorCouette2019} 
and \cite{martinez-ariasEffectNumberVortices2014}. These experiments employ different geometric parameters and significantly 
different aspect ratios, providing complementary validation across distinct parameter regimes. 
Figures \ref{fig:ramesh_comparison} and \ref{fig:martinez} show the comparisons 
with both experimental measurements.
Due to the smoothing function described in section \ref{sec:numerics}, we apply systematic corrections 
to match the experimental data. A $10\%$ shift in Reynolds number and a specific geometry dependent prefactor in the 
angular momentum flux were used for calibration such that the difference in angular momentum flux measurements in 
the laminar regime remains below $1\%$.

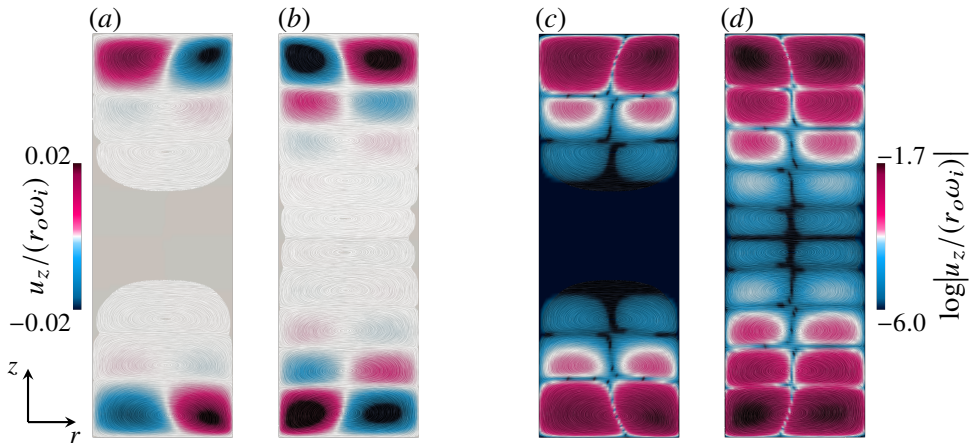
\begin{figure}[h!]
    \centering
    \hspace*{0.25cm}
    \begin{tikzpicture}
        \node[anchor=south west,inner sep=0] (image) at (0,0) 
            {\includegraphics[width=0.39\textwidth, trim=5cm 0 5cm 0, clip]{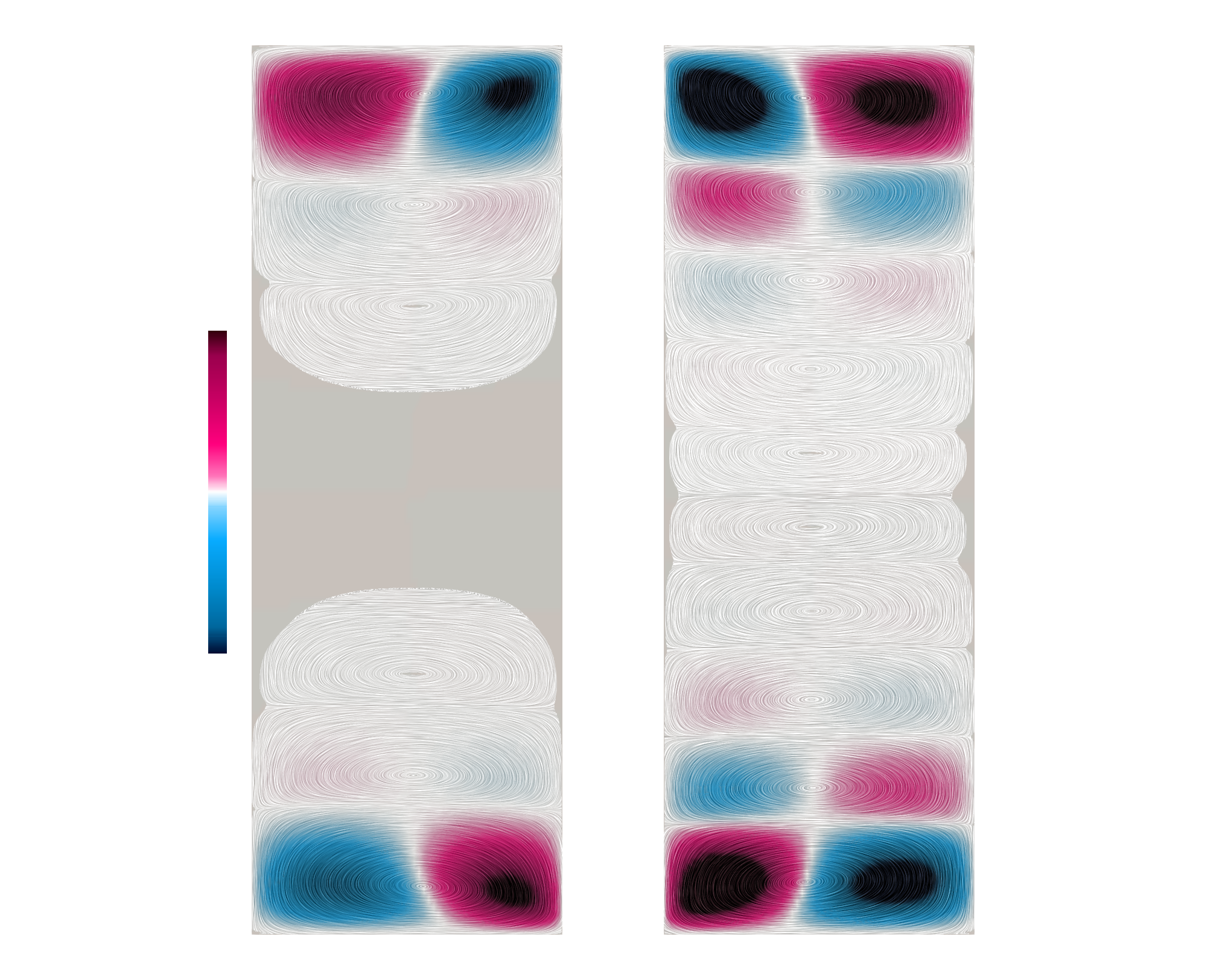}};
        \node[anchor=north west, yshift=0.2cm, xshift=0.3cm] at (image.north west) {(\textit{a})};
        \node[anchor=north west, yshift=0.2cm, xshift=2.8cm] at (image.north west) {(\textit{b})};
        
        \begin{scope}[x={(image.south east)},y={(image.north west)}]
            
            \node[scale=0.9] at (-0.04,0.68) {$\phantom{-}0.02$};
            
            \node[rotate=90] at (-0.05,0.5) {$\displaystyle u_z/(r_o\omega_i)$};
            
            \node[scale=0.9] at (-0.04,0.305) {$-0.02$};
            \draw[-stealth, thick] (-0.07,0.08) -- (0.05,0.08) node[below, scale=0.9] {$r$};
            \draw[-stealth, thick] (-0.07,0.08) -- (-0.07,0.2) node[left, scale=0.9] {$z$};
        \end{scope}
    \end{tikzpicture}
    \hspace{0.45cm}
    \begin{tikzpicture}
        \node[anchor=south west,inner sep=0] (image) at (0,0) 
            {\includegraphics[width=0.39\textwidth, trim=5cm 0 5cm 0, clip]{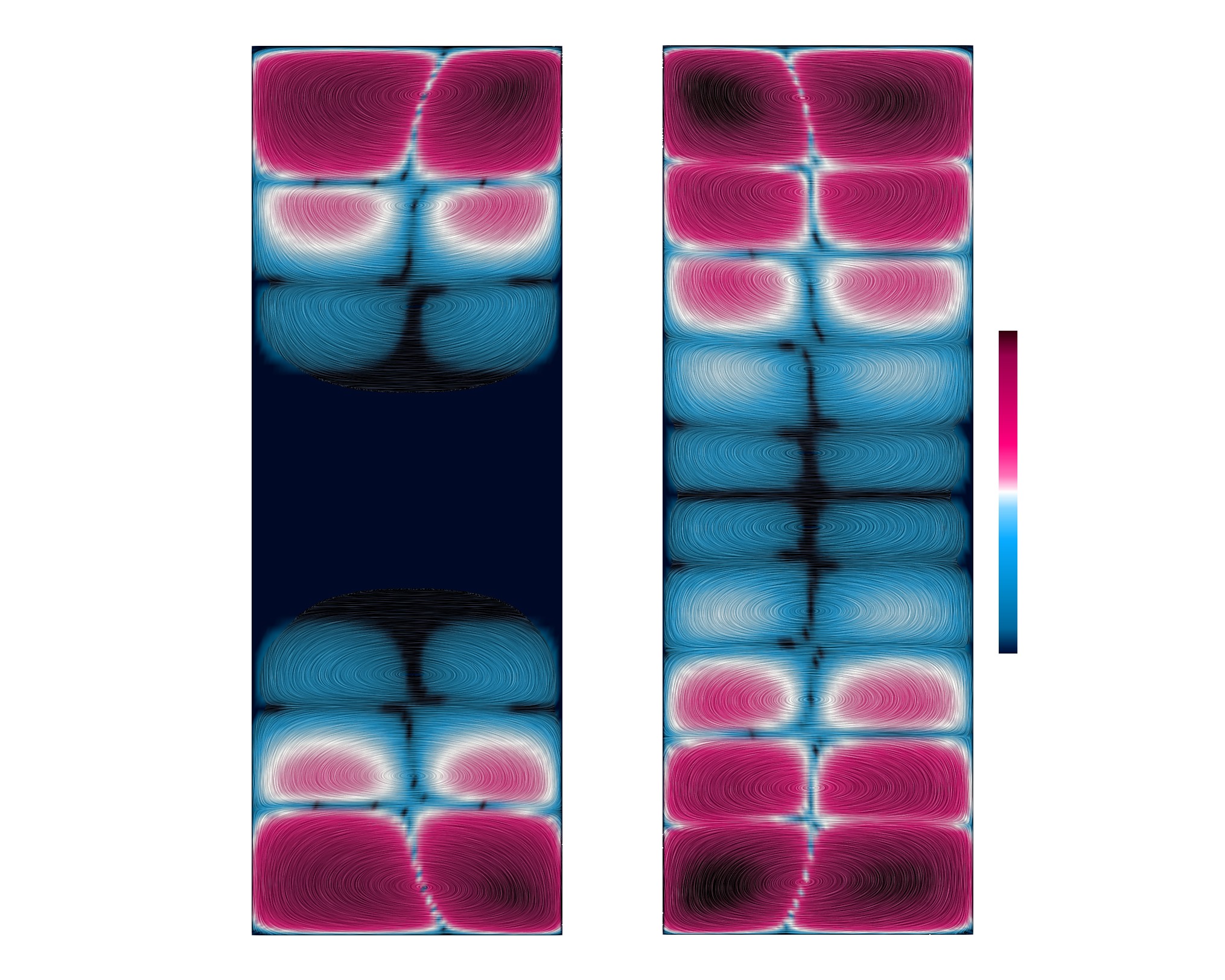}};
        \node[anchor=north west, yshift=0.2cm, xshift=0.3cm] at (image.north west) {(\textit{c})};
        \node[anchor=north west, yshift=0.2cm, xshift=2.75cm] at (image.north west) {(\textit{d})};
        
        \begin{scope}[x={(image.south east)},y={(image.north west)}]
            
            \node[scale=0.9] at (1,0.68) {$-1.7$};
            
            \node[rotate=90] at (1.125,0.5) {$\displaystyle\mathrm{log}\big|u_z/(r_o\omega_i)\big|$};
            
            \node[scale=0.9] at (1,0.305) {$-6.0$};
        \end{scope}
    \end{tikzpicture}
    \caption{Central vertical cross-sections of the instantaneous vertical velocity for two different Reynolds 
    numbers $(a,c)\,\,\Rey=90$ and $(b,d)\,\,\Rey=120$ in the setup of 
    \cite{rameshSuspensionTaylorCouette2019} for $\Gamma=11$ and $\eta=0.914$. The colour coding resolves the strength
    in axial velocity in $(\textit{a},\textit{b})$, where blue (red) denotes negative (positive) velocites. 
    Smooth contiguous regions indicate areas below the filter threshold of $10^{-6}$.
    In $(\textit{c},\textit{d})$ the amplitude of the axial velocity is shown on a logarithmic scale ranging from 
    lower (blue) to higher (red) velocities
    to enhance the contrast in roll-configuration visibility.
    $(\textit{a},\textit{c})$ At $\Rey=90$, convection rolls form initially from the upper and lower boundaries 
    and penetrate into the centre of the domain. Due to the limited angular frequency of the inner wall, 
    the roll formation cannot yet cover the full domain. $(\textit{b},\textit{d})$ At $\Rey=120$, 
    the external driving is strong enough to enable a domain-wide convection roll development.}
    \label{fig:flowfields_ramesh}
\end{figure}

We first compare our DNS results with the measurements of \cite{rameshSuspensionTaylorCouette2019} for $\eta = 0.914$ and $\Gamma = 11$. 
Figure \ref{fig:ramesh_comparison} demonstrates good agreement with the experimental results regarding 
both the onset of Taylor vortex flow and the subsequent angular momentum flux evolution.
The development of the flow field dynamics (depicted in figure \ref{fig:flowfields_ramesh}) supports the description 
by \cite{rameshSuspensionTaylorCouette2019} regarding the physical mechanisms responsible for the shift in the critical Reynolds 
number compared to theoretical estimates of $\Rey_c = 142.15$ \citep{diprima_swinney_1981} for infinitely long cylinders. The transition to TVF occurs at lower $\Rey$ than predicted 
by linear stability theory due to perturbations induced by Ekman vortices \citep{rameshSuspensionTaylorCouette2019}. 
Our flow visualizations clearly reveal the early self-organization of convection rolls in the vicinity of the 
lids, which progressively penetrate deeper into the flow domain as $\Rey$ increases.
However, a significant increase in the discrepancy between simulation and experiment can be observed at the 
upper end of the Reynolds number range. This deviation is most likely due to the formation of roll 
configurations that differ from those measured in the experiments and will be further discussed in 
chapter \ref{sec:multiple_states}.

\vspace{-0.5cm}

\newpage

\newpage
Another experimental study we compare with is by \cite{martinez-ariasEffectNumberVortices2014}
for a system with significantly different geometry $(\eta = 0.909$ and $\Gamma = 30)$.
Figure \ref{fig:martinez} shows a pronounced quantitative agreement between our DNS 
results (coloured symbols) and their experimental data (coloured lineplots) across the full range of $50 \leq \Rey \leq 5000$.
Our simulations with different initial roll numbers (18, 20, 22, 24, 26, and 28 rolls) all converge to the 
corresponding experimental curves in stable regimes, validating our numerical approach.

We also observe pronounced 
differences in flow dynamics depending on initial conditions. Simulations initialized with zero velocity 
fields and $\Rey\leq 4000$ consistently evolve toward conditionally stable Taylor vortex flow states with 
asymmetric roll configurations. However, higher Reynolds numbers impose sufficient disturbances to drive 
the system out of the Taylor-Vortex attractor towards more wavy or modulated vortex states. 
By adding white noise perturbations to the initial velocity field, we observed significant dynamical shifts. 
Specifically, systems underwent transitions from TVF to WVF in regimes where WVF has been observed experimentally. 
This observation motivated the initialization 
procedure described in section \ref{seq:appendix}, whereby we initialize the system with disturbed Taylor vortex flow of the 
desired roll number. This approach enabled us to systematically study the occurrence and dynamics of multiple 
states while maintaining excellent agreement with experimental measurements (for detailed discussion on 
multiple states see section \ref{sec:multiple_states}).

\subsection{Boundary layer theory with axial corrections}

\begin{figure}[h!]
    \centering
    \includegraphics{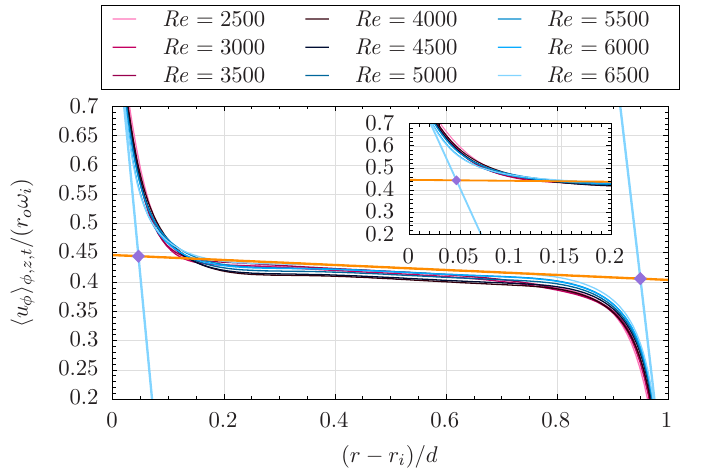}
    \vspace{-0.2cm}
    \caption{Sketch of the boundary layer thickness estimation for $2.5\times 10^3 \leq \Rey \leq 6.5 \times 10^3$ in the 
    setup of \cite{martinez-ariasEffectNumberVortices2014} for $\Gamma=30$ and $\eta=0.909$. 
    Violet diamonds denote the estimated width of the boundary layer obtained via the slope method: 
    linear fits to the two wall-most points are extrapolated to intersect a horizontal line at the bulk angular velocity.}
    \label{fig:boundary_layer_1}
\end{figure}

To examine the spatial structure in detail, figure \ref{fig:boundary_layer_1} shows
radial profiles of the averaged angular velocity $\langle u_\phi\rangle_{\phi,z,t}$ for different Reynolds 
numbers. At low $\Rey$, the monotonic profiles exhibit 
clear boundary layers near both cylinders with a weakly changing bulk region. At higher $\Rey$, the steeper gradients 
near the walls reflect the increased importance of turbulent fluctuations in momentum transport.

\begin{figure}[h!]
    \centering
    \includegraphics{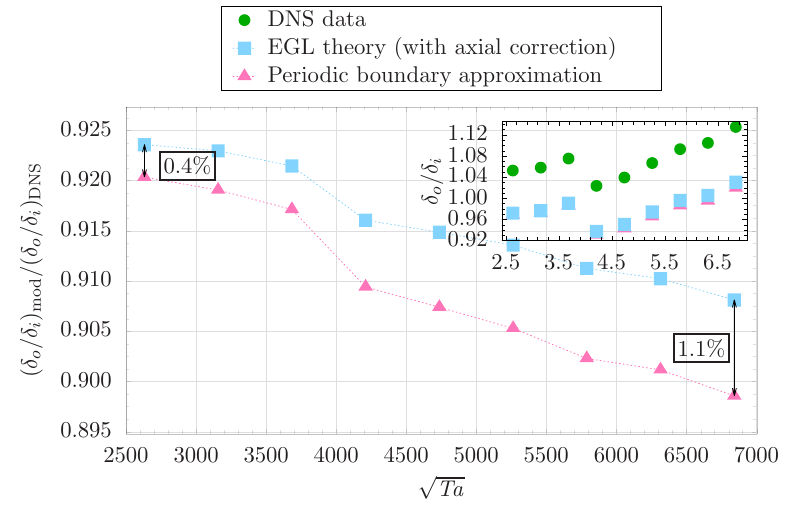}
    \caption{Ratio of the boundary layer thicknesses based on theoretical predictions compared to the ones obtained 
    in the DNS for $2.5\times10^3 < \sqrt{\mathit{Ta}} < 7 \times 10^3$ in the setup of 
    \cite{martinez-ariasEffectNumberVortices2014} for $\Gamma=30$ and $\eta=0.909$.
    Shown is the model predicted ratio (\ref{eq:bl_thickness}) against the DNS results estimated based on the 
    procedure shown in figure \ref{fig:boundary_layer_1}. 
    The inset in the top right corner shows the non-normalized ratios against $\sqrt{\mathit{Ta}}\times 10^{-3}$. 
    The deviation between modified theory and periodic approximation grows significantly with higher $\mathit{Ta}$.}
    \label{fig:boundary_layer}
\end{figure}

\noindent Using the extended angular momentum flux formulation validated above, we now examine boundary layer structure 
and compare our DNS results with theoretical predictions.
A key quantity for characterizing flow asymmetry is the ratio of outer to inner boundary layer 
thicknesses, $\delta_{o}/\delta_{i}$. Figure \ref{fig:boundary_layer}
presents this ratio as a function of $\sqrt{\mathit{Ta}}$.
The DNS data are compared with the extended theory including axial corrections (cyan squares in 
figure \ref{fig:boundary_layer}) as introduced 
in section \ref{sec:system}, and the periodic boundary approximation (magenta triangles)
representing \cite{eckhardtTorqueScalingTurbulent2007} predictions that neglect axial transport.
Figure \ref{fig:boundary_layer} reveals several important features. 
As shown in figure \ref{fig:martinez}, simulations at $\Rey < 4000$ consistently evolve toward a stable TVF state. 
Within this laminar to weakly nonlinear regime, the boundary-layer thickness ratio $\delta_o/\delta_i$ 
exceeds unity and grows with increasing $\mathit{Ta}$,  
reflecting the stronger mean shear at the rotating inner cylinder relative to the stationary outer wall.
Upon transition to WVF, the ratio drops significantly. This behavior is attributed to coherent wave structures that 
preferentially enhance angular momentum transport toward the outer cylinder. 
Consequently, the outer boundary layer thins significantly, while the inner boundary layer, which is already governed 
by the imposed rotational shear, remains largely unaffected. 
This results in a reduction of the thickness ratio.
The predicted boundary layer ratio based on equation (\ref{eq:bl_thickness}) consistently underestimates the DNS results. 
This underestimation also increases with higher $\mathit{Ta}$ due to steeper radial gradients of $\omega$, which 
are not sufficiently captured by our linear approximation (\ref{eq:bc_thick}).

Interestingly, we also observe a significant increase in the deviation between the periodic approximation and the 
modified model for higher rotation rates of the inner cylinder. 
This finding is consistent with our angular momentum flux analysis shown in section \ref{seq:appendix_angular_flux}, 
highlighting the dependence of endwall boundary layer effects as a function of $\mathit{Ta}$.
The corrections introduced by the additional term in equation (\ref{eq:bl_thickness}) are therefore quantitatively 
significant and cannot be neglected for accurate predictions in realistic geometries.

\section{Multiple stable states and hysteresis}
\label{sec:multiple_states}

To systematically explore the stability landscape, we performed DNS runs at Reynolds numbers $\Rey \in \{400, 3000\}$, in the 
setup of $\Gamma = 30$ and $\eta = 0.909$,
with initial conditions covering different roll number configurations $n \in \{18, 20, 22, 24, 26, 28\}$, based on the 
experimental findings by \cite{martinez-ariasEffectNumberVortices2014}. 
Each simulation was run for sufficient time to determine the final persistent state, typically exceeding 600 rotational time
units $(t_{\text{rot}}=t \omega_i)$.
Our simulations reveal that multiple distinct stable states can coexist at identical Reynolds
numbers, with the final roll number depending sensitively on the initial condition.
However, we also observe the occurrence of additional stable states not reported in their
experimental work. This discrepancy likely arises because numerical simulations are
significantly less susceptible to external perturbations than experimental investigations.
It should be noted, however, that some configurations may require longer integration times
to fully capture potential transitions to lower roll numbers.
\vspace{0.2cm}
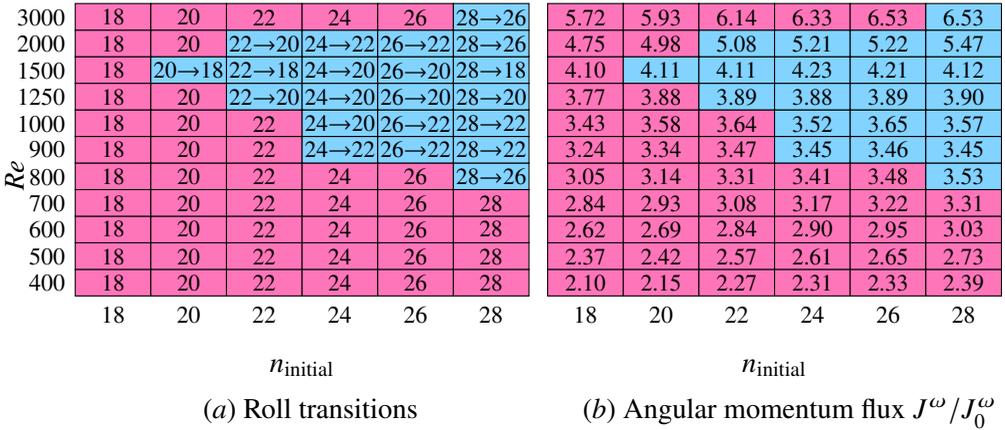
\begin{figure}[h!]
    \centering
    \hspace*{-0.5cm}
    \begin{minipage}[b]{0.45\textwidth}
        \centering
        \begin{tikzpicture}[xscale=1.0, yscale=1.0]  

        \fill[colorstable] (0.00,0.00) rectangle (1.00,0.35); \draw[black,thin] (0.00,0.00) rectangle (1.00,0.35); \node[font=\footnotesize,black] at (0.50,0.17) {18};
        \fill[colorstable] (1.00,0.00) rectangle (2.00,0.35); \draw[black,thin] (1.00,0.00) rectangle (2.00,0.35); \node[font=\footnotesize,black] at (1.50,0.17) {20};
        \fill[colorstable] (2.00,0.00) rectangle (3.00,0.35); \draw[black,thin] (2.00,0.00) rectangle (3.00,0.35); \node[font=\footnotesize,black] at (2.50,0.17) {22};
        \fill[colorstable] (3.00,0.00) rectangle (4.00,0.35); \draw[black,thin] (3.00,0.00) rectangle (4.00,0.35); \node[font=\footnotesize,black] at (3.50,0.17) {24};
        \fill[colorstable] (4.00,0.00) rectangle (5.00,0.35); \draw[black,thin] (4.00,0.00) rectangle (5.00,0.35); \node[font=\footnotesize,black] at (4.50,0.17) {26};
        \fill[colorstable] (5.00,0.00) rectangle (6.00,0.35); \draw[black,thin] (5.00,0.00) rectangle (6.00,0.35); \node[font=\footnotesize,black] at (5.50,0.17) {28};

        \fill[colorstable] (0.00,0.35) rectangle (1.00,0.70); \draw[black,thin] (0.00,0.35) rectangle (1.00,0.70); \node[font=\footnotesize,black] at (0.50,0.52) {18};
        \fill[colorstable] (1.00,0.35) rectangle (2.00,0.70); \draw[black,thin] (1.00,0.35) rectangle (2.00,0.70); \node[font=\footnotesize,black] at (1.50,0.52) {20};
        \fill[colorstable] (2.00,0.35) rectangle (3.00,0.70); \draw[black,thin] (2.00,0.35) rectangle (3.00,0.70); \node[font=\footnotesize,black] at (2.50,0.52) {22};
        \fill[colorstable] (3.00,0.35) rectangle (4.00,0.70); \draw[black,thin] (3.00,0.35) rectangle (4.00,0.70); \node[font=\footnotesize,black] at (3.50,0.52) {24};
        \fill[colorstable] (4.00,0.35) rectangle (5.00,0.70); \draw[black,thin] (4.00,0.35) rectangle (5.00,0.70); \node[font=\footnotesize,black] at (4.50,0.52) {26};
        \fill[colorstable] (5.00,0.35) rectangle (6.00,0.70); \draw[black,thin] (5.00,0.35) rectangle (6.00,0.70); \node[font=\footnotesize,black] at (5.50,0.52) {28};

        \fill[colorstable] (0.00,0.70) rectangle (1.00,1.05); \draw[black,thin] (0.00,0.70) rectangle (1.00,1.05); \node[font=\footnotesize,black] at (0.50,0.88) {18};
        \fill[colorstable] (1.00,0.70) rectangle (2.00,1.05); \draw[black,thin] (1.00,0.70) rectangle (2.00,1.05); \node[font=\footnotesize,black] at (1.50,0.88) {20};
        \fill[colorstable] (2.00,0.70) rectangle (3.00,1.05); \draw[black,thin] (2.00,0.70) rectangle (3.00,1.05); \node[font=\footnotesize,black] at (2.50,0.88) {22};
        \fill[colorstable] (3.00,0.70) rectangle (4.00,1.05); \draw[black,thin] (3.00,0.70) rectangle (4.00,1.05); \node[font=\footnotesize,black] at (3.50,0.88) {24};
        \fill[colorstable] (4.00,0.70) rectangle (5.00,1.05); \draw[black,thin] (4.00,0.70) rectangle (5.00,1.05); \node[font=\footnotesize,black] at (4.50,0.88) {26};
        \fill[colorstable] (5.00,0.70) rectangle (6.00,1.05); \draw[black,thin] (5.00,0.70) rectangle (6.00,1.05); \node[font=\footnotesize,black] at (5.50,0.88) {28};

        \fill[colorstable] (0.00,1.05) rectangle (1.00,1.40); \draw[black,thin] (0.00,1.05) rectangle (1.00,1.40); \node[font=\footnotesize,black] at (0.50,1.22) {18};
        \fill[colorstable] (1.00,1.05) rectangle (2.00,1.40); \draw[black,thin] (1.00,1.05) rectangle (2.00,1.40); \node[font=\footnotesize,black] at (1.50,1.22) {20};
        \fill[colorstable] (2.00,1.05) rectangle (3.00,1.40); \draw[black,thin] (2.00,1.05) rectangle (3.00,1.40); \node[font=\footnotesize,black] at (2.50,1.22) {22};
        \fill[colorstable] (3.00,1.05) rectangle (4.00,1.40); \draw[black,thin] (3.00,1.05) rectangle (4.00,1.40); \node[font=\footnotesize,black] at (3.50,1.22) {24};
        \fill[colorstable] (4.00,1.05) rectangle (5.00,1.40); \draw[black,thin] (4.00,1.05) rectangle (5.00,1.40); \node[font=\footnotesize,black] at (4.50,1.22) {26};
        \fill[colorstable] (5.00,1.05) rectangle (6.00,1.40); \draw[black,thin] (5.00,1.05) rectangle (6.00,1.40); \node[font=\footnotesize,black] at (5.50,1.22) {28};

        \fill[colorstable] (0.00,1.40) rectangle (1.00,1.75); \draw[black,thin] (0.00,1.40) rectangle (1.00,1.75); \node[font=\footnotesize,black] at (0.50,1.57) {18};
        \fill[colorstable] (1.00,1.40) rectangle (2.00,1.75); \draw[black,thin] (1.00,1.40) rectangle (2.00,1.75); \node[font=\footnotesize,black] at (1.50,1.57) {20};
        \fill[colorstable] (2.00,1.40) rectangle (3.00,1.75); \draw[black,thin] (2.00,1.40) rectangle (3.00,1.75); \node[font=\footnotesize,black] at (2.50,1.57) {22};
        \fill[colorstable] (3.00,1.40) rectangle (4.00,1.75); \draw[black,thin] (3.00,1.40) rectangle (4.00,1.75); \node[font=\footnotesize,black] at (3.50,1.57) {24};
        \fill[colorstable] (4.00,1.40) rectangle (5.00,1.75); \draw[black,thin] (4.00,1.40) rectangle (5.00,1.75); \node[font=\footnotesize,black] at (4.50,1.57) {26};
        \fill[colorunstable] (5.00,1.40) rectangle (6.00,1.75); \draw[black,thin] (5.00,1.40) rectangle (6.00,1.75); \node[font=\footnotesize,black] at (5.50,1.57) {28$\rightarrow$26};
 
        \fill[colorstable] (0.00,1.75) rectangle (1.00,2.10); \draw[black,thin] (0.00,1.75) rectangle (1.00,2.10); \node[font=\footnotesize,black] at (0.50,1.93) {18};
        \fill[colorstable] (1.00,1.75) rectangle (2.00,2.10); \draw[black,thin] (1.00,1.75) rectangle (2.00,2.10); \node[font=\footnotesize,black] at (1.50,1.93) {20};
        \fill[colorstable] (2.00,1.75) rectangle (3.00,2.10); \draw[black,thin] (2.00,1.75) rectangle (3.00,2.10); \node[font=\footnotesize,black] at (2.50,1.93) {22};
        \fill[colorunstable] (3.00,1.75) rectangle (4.00,2.10); \draw[black,thin] (3.00,1.75) rectangle (4.00,2.10); \node[font=\footnotesize,black] at (3.50,1.93) {24$\rightarrow$22};
        \fill[colorunstable] (4.00,1.75) rectangle (5.00,2.10); \draw[black,thin] (4.00,1.75) rectangle (5.00,2.10); \node[font=\footnotesize,black] at (4.50,1.93) {26$\rightarrow$22};
        \fill[colorunstable] (5.00,1.75) rectangle (6.00,2.10); \draw[black,thin] (5.00,1.75) rectangle (6.00,2.10); \node[font=\footnotesize,black] at (5.50,1.93) {28$\rightarrow$22};

        \fill[colorstable] (0.00,2.10) rectangle (1.00,2.45); \draw[black,thin] (0.00,2.10) rectangle (1.00,2.45); \node[font=\footnotesize,black] at (0.50,2.27) {18};
        \fill[colorstable] (1.00,2.10) rectangle (2.00,2.45); \draw[black,thin] (1.00,2.10) rectangle (2.00,2.45); \node[font=\footnotesize,black] at (1.50,2.27) {20};
        \fill[colorstable] (2.00,2.10) rectangle (3.00,2.45); \draw[black,thin] (2.00,2.10) rectangle (3.00,2.45); \node[font=\footnotesize,black] at (2.50,2.27) {22};
        \fill[colorunstable] (3.00,2.10) rectangle (4.00,2.45); \draw[black,thin] (3.00,2.10) rectangle (4.00,2.45); \node[font=\footnotesize,black] at (3.50,2.27) {24$\rightarrow$20};
        \fill[colorunstable] (4.00,2.10) rectangle (5.00,2.45); \draw[black,thin] (4.00,2.10) rectangle (5.00,2.45); \node[font=\footnotesize,black] at (4.50,2.27) {26$\rightarrow$22};
        \fill[colorunstable] (5.00,2.10) rectangle (6.00,2.45); \draw[black,thin] (5.00,2.10) rectangle (6.00,2.45); \node[font=\footnotesize,black] at (5.50,2.27) {28$\rightarrow$22};

        \fill[colorstable] (0.00,2.45) rectangle (1.00,2.80); \draw[black,thin] (0.00,2.45) rectangle (1.00,2.80); \node[font=\footnotesize,black] at (0.50,2.62) {18};
        \fill[colorstable] (1.00,2.45) rectangle (2.00,2.80); \draw[black,thin] (1.00,2.45) rectangle (2.00,2.80); \node[font=\footnotesize,black] at (1.50,2.62) {20};
        \fill[colorunstable] (2.00,2.45) rectangle (3.00,2.80); \draw[black,thin] (2.00,2.45) rectangle (3.00,2.80); \node[font=\footnotesize,black] at (2.50,2.62) {22$\rightarrow$20};
        \fill[colorunstable] (3.00,2.45) rectangle (4.00,2.80); \draw[black,thin] (3.00,2.45) rectangle (4.00,2.80); \node[font=\footnotesize,black] at (3.50,2.62) {24$\rightarrow$20};
        \fill[colorunstable] (4.00,2.45) rectangle (5.00,2.80); \draw[black,thin] (4.00,2.45) rectangle (5.00,2.80); \node[font=\footnotesize,black] at (4.50,2.62) {26$\rightarrow$20};
        \fill[colorunstable] (5.00,2.45) rectangle (6.00,2.80); \draw[black,thin] (5.00,2.45) rectangle (6.00,2.80); \node[font=\footnotesize,black] at (5.50,2.62) {28$\rightarrow$20};

        \fill[colorstable] (0.00,2.80) rectangle (1.00,3.15); \draw[black,thin] (0.00,2.80) rectangle (1.00,3.15); \node[font=\footnotesize,black] at (0.50,2.97) {18};
        \fill[colorunstable] (1.00,2.80) rectangle (2.00,3.15); \draw[black,thin] (1.00,2.80) rectangle (2.00,3.15); \node[font=\footnotesize,black] at (1.50,2.97) {20$\rightarrow$18};
        \fill[colorunstable] (2.00,2.80) rectangle (3.00,3.15); \draw[black,thin] (2.00,2.80) rectangle (3.00,3.15); \node[font=\footnotesize,black] at (2.50,2.97) {22$\rightarrow$18};
        \fill[colorunstable] (3.00,2.80) rectangle (4.00,3.15); \draw[black,thin] (3.00,2.80) rectangle (4.00,3.15); \node[font=\footnotesize,black] at (3.50,2.97) {24$\rightarrow$20};
        \fill[colorunstable] (4.00,2.80) rectangle (5.00,3.15); \draw[black,thin] (4.00,2.80) rectangle (5.00,3.15); \node[font=\footnotesize,black] at (4.50,2.97) {26$\rightarrow$20};
        \fill[colorunstable] (5.00,2.80) rectangle (6.00,3.15); \draw[black,thin] (5.00,2.80) rectangle (6.00,3.15); \node[font=\footnotesize,black] at (5.50,2.97) {28$\rightarrow$18};

        \fill[colorstable] (0.00,3.15) rectangle (1.00,3.50); \draw[black,thin] (0.00,3.15) rectangle (1.00,3.50); \node[font=\footnotesize,black] at (0.50,3.32) {18};
        \fill[colorstable] (1.00,3.15) rectangle (2.00,3.50); \draw[black,thin] (1.00,3.15) rectangle (2.00,3.50); \node[font=\footnotesize,black] at (1.50,3.32) {20};
        \fill[colorunstable] (2.00,3.15) rectangle (3.00,3.50); \draw[black,thin] (2.00,3.15) rectangle (3.00,3.50); \node[font=\footnotesize,black] at (2.50,3.32) {22$\rightarrow$20};
        \fill[colorunstable] (3.00,3.15) rectangle (4.00,3.50); \draw[black,thin] (3.00,3.15) rectangle (4.00,3.50); \node[font=\footnotesize,black] at (3.50,3.32) {24$\rightarrow$22};
        \fill[colorunstable] (4.00,3.15) rectangle (5.00,3.50); \draw[black,thin] (4.00,3.15) rectangle (5.00,3.50); \node[font=\footnotesize,black] at (4.50,3.32) {26$\rightarrow$22};
        \fill[colorunstable] (5.00,3.15) rectangle (6.00,3.50); \draw[black,thin] (5.00,3.15) rectangle (6.00,3.50); \node[font=\footnotesize,black] at (5.50,3.32) {28$\rightarrow$26};

        \fill[colorstable] (0.00,3.50) rectangle (1.00,3.85); \draw[black,thin] (0.00,3.50) rectangle (1.00,3.85); \node[font=\footnotesize,black] at (0.50,3.67) {18};
        \fill[colorstable] (1.00,3.50) rectangle (2.00,3.85); \draw[black,thin] (1.00,3.50) rectangle (2.00,3.85); \node[font=\footnotesize,black] at (1.50,3.67) {20};
        \fill[colorstable] (2.00,3.50) rectangle (3.00,3.85); \draw[black,thin] (2.00,3.50) rectangle (3.00,3.85); \node[font=\footnotesize,black] at (2.50,3.67) {22};
        \fill[colorstable] (3.00,3.50) rectangle (4.00,3.85); \draw[black,thin] (3.00,3.50) rectangle (4.00,3.85); \node[font=\footnotesize,black] at (3.50,3.67) {24};
        \fill[colorstable] (4.00,3.50) rectangle (5.00,3.85); \draw[black,thin] (4.00,3.50) rectangle (5.00,3.85); \node[font=\footnotesize,black] at (4.50,3.67) {26};
        \fill[colorunstable] (5.00,3.50) rectangle (6.00,3.85); \draw[black,thin] (5.00,3.50) rectangle (6.00,3.85); \node[font=\footnotesize,black] at (5.50,3.67) {28$\rightarrow$26};

        \node[below,font=\small] at (0.50,0) {18};
        \node[below,font=\small] at (1.50,0) {20};
        \node[below,font=\small] at (2.50,0) {22};
        \node[below,font=\small] at (3.50,0) {24};
        \node[below,font=\small] at (4.50,0) {26};
        \node[below,font=\small] at (5.50,0) {28};
        \node[below,font=\normalsize] at (3.00,-0.7) {$n_{\mathrm{initial}}$};

        \node[left,font=\small] at (0,0.17) {400};
        \node[left,font=\small] at (0,0.52) {500};
        \node[left,font=\small] at (0,0.88) {600};
        \node[left,font=\small] at (0,1.22) {700};
        \node[left,font=\small] at (0,1.57) {800};
        \node[left,font=\small] at (0,1.93) {900};
        \node[left,font=\small] at (0,2.27) {1000};
        \node[left,font=\small] at (0,2.62) {1250};
        \node[left,font=\small] at (0,2.97) {1500};
        \node[left,font=\small] at (0,3.32) {2000};
        \node[left,font=\small] at (0,3.67) {3000};
        \node[left,font=\normalsize,rotate=90] at (-0.8,1.92) {$\mathit{Re}$};

        \end{tikzpicture}
        
        \hspace{2.25cm}(\textit{a}) Roll transitions
    \end{minipage}
    \hspace{1.0cm}
    \begin{minipage}[b]{0.45\textwidth}
        \centering
        \begin{tikzpicture}[xscale=1.0, yscale=1.0] 

        \fill[colorstable] (0.00,0.00) rectangle (1.00,0.35); \draw[black,thin] (0.00,0.00) rectangle (1.00,0.35); \node[font=\footnotesize,black] at (0.50,0.17) {2.10};
        \fill[colorstable] (1.00,0.00) rectangle (2.00,0.35); \draw[black,thin] (1.00,0.00) rectangle (2.00,0.35); \node[font=\footnotesize,black] at (1.50,0.17) {2.15};
        \fill[colorstable] (2.00,0.00) rectangle (3.00,0.35); \draw[black,thin] (2.00,0.00) rectangle (3.00,0.35); \node[font=\footnotesize,black] at (2.50,0.17) {2.27};
        \fill[colorstable] (3.00,0.00) rectangle (4.00,0.35); \draw[black,thin] (3.00,0.00) rectangle (4.00,0.35); \node[font=\footnotesize,black] at (3.50,0.17) {2.31};
        \fill[colorstable] (4.00,0.00) rectangle (5.00,0.35); \draw[black,thin] (4.00,0.00) rectangle (5.00,0.35); \node[font=\footnotesize,black] at (4.50,0.17) {2.33};
        \fill[colorstable] (5.00,0.00) rectangle (6.00,0.35); \draw[black,thin] (5.00,0.00) rectangle (6.00,0.35); \node[font=\footnotesize,black] at (5.50,0.17) {2.39};

        \fill[colorstable] (0.00,0.35) rectangle (1.00,0.70); \draw[black,thin] (0.00,0.35) rectangle (1.00,0.70); \node[font=\footnotesize,black] at (0.50,0.52) {2.37};
        \fill[colorstable] (1.00,0.35) rectangle (2.00,0.70); \draw[black,thin] (1.00,0.35) rectangle (2.00,0.70); \node[font=\footnotesize,black] at (1.50,0.52) {2.42};
        \fill[colorstable] (2.00,0.35) rectangle (3.00,0.70); \draw[black,thin] (2.00,0.35) rectangle (3.00,0.70); \node[font=\footnotesize,black] at (2.50,0.52) {2.57};
        \fill[colorstable] (3.00,0.35) rectangle (4.00,0.70); \draw[black,thin] (3.00,0.35) rectangle (4.00,0.70); \node[font=\footnotesize,black] at (3.50,0.52) {2.61};
        \fill[colorstable] (4.00,0.35) rectangle (5.00,0.70); \draw[black,thin] (4.00,0.35) rectangle (5.00,0.70); \node[font=\footnotesize,black] at (4.50,0.52) {2.65};
        \fill[colorstable] (5.00,0.35) rectangle (6.00,0.70); \draw[black,thin] (5.00,0.35) rectangle (6.00,0.70); \node[font=\footnotesize,black] at (5.50,0.52) {2.73};

        \fill[colorstable] (0.00,0.70) rectangle (1.00,1.05); \draw[black,thin] (0.00,0.70) rectangle (1.00,1.05); \node[font=\footnotesize,black] at (0.50,0.88) {2.62};
        \fill[colorstable] (1.00,0.70) rectangle (2.00,1.05); \draw[black,thin] (1.00,0.70) rectangle (2.00,1.05); \node[font=\footnotesize,black] at (1.50,0.88) {2.69};
        \fill[colorstable] (2.00,0.70) rectangle (3.00,1.05); \draw[black,thin] (2.00,0.70) rectangle (3.00,1.05); \node[font=\footnotesize,black] at (2.50,0.88) {2.84};
        \fill[colorstable] (3.00,0.70) rectangle (4.00,1.05); \draw[black,thin] (3.00,0.70) rectangle (4.00,1.05); \node[font=\footnotesize,black] at (3.50,0.88) {2.90};
        \fill[colorstable] (4.00,0.70) rectangle (5.00,1.05); \draw[black,thin] (4.00,0.70) rectangle (5.00,1.05); \node[font=\footnotesize,black] at (4.50,0.88) {2.95};
        \fill[colorstable] (5.00,0.70) rectangle (6.00,1.05); \draw[black,thin] (5.00,0.70) rectangle (6.00,1.05); \node[font=\footnotesize,black] at (5.50,0.88) {3.03};

        \fill[colorstable] (0.00,1.05) rectangle (1.00,1.40); \draw[black,thin] (0.00,1.05) rectangle (1.00,1.40); \node[font=\footnotesize,black] at (0.50,1.22) {2.84};
        \fill[colorstable] (1.00,1.05) rectangle (2.00,1.40); \draw[black,thin] (1.00,1.05) rectangle (2.00,1.40); \node[font=\footnotesize,black] at (1.50,1.22) {2.93};
        \fill[colorstable] (2.00,1.05) rectangle (3.00,1.40); \draw[black,thin] (2.00,1.05) rectangle (3.00,1.40); \node[font=\footnotesize,black] at (2.50,1.22) {3.08};
        \fill[colorstable] (3.00,1.05) rectangle (4.00,1.40); \draw[black,thin] (3.00,1.05) rectangle (4.00,1.40); \node[font=\footnotesize,black] at (3.50,1.22) {3.17};
        \fill[colorstable] (4.00,1.05) rectangle (5.00,1.40); \draw[black,thin] (4.00,1.05) rectangle (5.00,1.40); \node[font=\footnotesize,black] at (4.50,1.22) {3.22};
        \fill[colorstable] (5.00,1.05) rectangle (6.00,1.40); \draw[black,thin] (5.00,1.05) rectangle (6.00,1.40); \node[font=\footnotesize,black] at (5.50,1.22) {3.31};

        \fill[colorstable] (0.00,1.40) rectangle (1.00,1.75); \draw[black,thin] (0.00,1.40) rectangle (1.00,1.75); \node[font=\footnotesize,black] at (0.50,1.57) {3.05};
        \fill[colorstable] (1.00,1.40) rectangle (2.00,1.75); \draw[black,thin] (1.00,1.40) rectangle (2.00,1.75); \node[font=\footnotesize,black] at (1.50,1.57) {3.14};
        \fill[colorstable] (2.00,1.40) rectangle (3.00,1.75); \draw[black,thin] (2.00,1.40) rectangle (3.00,1.75); \node[font=\footnotesize,black] at (2.50,1.57) {3.31};
        \fill[colorstable] (3.00,1.40) rectangle (4.00,1.75); \draw[black,thin] (3.00,1.40) rectangle (4.00,1.75); \node[font=\footnotesize,black] at (3.50,1.57) {3.41};
        \fill[colorstable] (4.00,1.40) rectangle (5.00,1.75); \draw[black,thin] (4.00,1.40) rectangle (5.00,1.75); \node[font=\footnotesize,black] at (4.50,1.57) {3.48};
        \fill[colorunstable] (5.00,1.40) rectangle (6.00,1.75); \draw[black,thin] (5.00,1.40) rectangle (6.00,1.75); \node[font=\footnotesize,black] at (5.50,1.57) {3.53};

        \fill[colorstable] (0.00,1.75) rectangle (1.00,2.10); \draw[black,thin] (0.00,1.75) rectangle (1.00,2.10); \node[font=\footnotesize,black] at (0.50,1.93) {3.24};
        \fill[colorstable] (1.00,1.75) rectangle (2.00,2.10); \draw[black,thin] (1.00,1.75) rectangle (2.00,2.10); \node[font=\footnotesize,black] at (1.50,1.93) {3.34};
        \fill[colorstable] (2.00,1.75) rectangle (3.00,2.10); \draw[black,thin] (2.00,1.75) rectangle (3.00,2.10); \node[font=\footnotesize,black] at (2.50,1.93) {3.47};
        \fill[colorunstable] (3.00,1.75) rectangle (4.00,2.10); \draw[black,thin] (3.00,1.75) rectangle (4.00,2.10); \node[font=\footnotesize,black] at (3.50,1.93) {3.45};
        \fill[colorunstable] (4.00,1.75) rectangle (5.00,2.10); \draw[black,thin] (4.00,1.75) rectangle (5.00,2.10); \node[font=\footnotesize,black] at (4.50,1.93) {3.46};
        \fill[colorunstable] (5.00,1.75) rectangle (6.00,2.10); \draw[black,thin] (5.00,1.75) rectangle (6.00,2.10); \node[font=\footnotesize,black] at (5.50,1.93) {3.45};

        \fill[colorstable] (0.00,2.10) rectangle (1.00,2.45); \draw[black,thin] (0.00,2.10) rectangle (1.00,2.45); \node[font=\footnotesize,black] at (0.50,2.27) {3.43};
        \fill[colorstable] (1.00,2.10) rectangle (2.00,2.45); \draw[black,thin] (1.00,2.10) rectangle (2.00,2.45); \node[font=\footnotesize,black] at (1.50,2.27) {3.58};
        \fill[colorstable] (2.00,2.10) rectangle (3.00,2.45); \draw[black,thin] (2.00,2.10) rectangle (3.00,2.45); \node[font=\footnotesize,black] at (2.50,2.27) {3.64};
        \fill[colorunstable] (3.00,2.10) rectangle (4.00,2.45); \draw[black,thin] (3.00,2.10) rectangle (4.00,2.45); \node[font=\footnotesize,black] at (3.50,2.27) {3.52};
        \fill[colorunstable] (4.00,2.10) rectangle (5.00,2.45); \draw[black,thin] (4.00,2.10) rectangle (5.00,2.45); \node[font=\footnotesize,black] at (4.50,2.27) {3.65};
        \fill[colorunstable] (5.00,2.10) rectangle (6.00,2.45); \draw[black,thin] (5.00,2.10) rectangle (6.00,2.45); \node[font=\footnotesize,black] at (5.50,2.27) {3.57};

        \fill[colorstable] (0.00,2.45) rectangle (1.00,2.80); \draw[black,thin] (0.00,2.45) rectangle (1.00,2.80); \node[font=\footnotesize,black] at (0.50,2.62) {3.77};
        \fill[colorstable] (1.00,2.45) rectangle (2.00,2.80); \draw[black,thin] (1.00,2.45) rectangle (2.00,2.80); \node[font=\footnotesize,black] at (1.50,2.62) {3.88};
        \fill[colorunstable] (2.00,2.45) rectangle (3.00,2.80); \draw[black,thin] (2.00,2.45) rectangle (3.00,2.80); \node[font=\footnotesize,black] at (2.50,2.62) {3.89};
        \fill[colorunstable] (3.00,2.45) rectangle (4.00,2.80); \draw[black,thin] (3.00,2.45) rectangle (4.00,2.80); \node[font=\footnotesize,black] at (3.50,2.62) {3.88};
        \fill[colorunstable] (4.00,2.45) rectangle (5.00,2.80); \draw[black,thin] (4.00,2.45) rectangle (5.00,2.80); \node[font=\footnotesize,black] at (4.50,2.62) {3.89};
        \fill[colorunstable] (5.00,2.45) rectangle (6.00,2.80); \draw[black,thin] (5.00,2.45) rectangle (6.00,2.80); \node[font=\footnotesize,black] at (5.50,2.62) {3.90};

        \fill[colorstable] (0.00,2.80) rectangle (1.00,3.15); \draw[black,thin] (0.00,2.80) rectangle (1.00,3.15); \node[font=\footnotesize,black] at (0.50,2.97) {4.10};
        \fill[colorunstable] (1.00,2.80) rectangle (2.00,3.15); \draw[black,thin] (1.00,2.80) rectangle (2.00,3.15); \node[font=\footnotesize,black] at (1.50,2.97) {4.11};
        \fill[colorunstable] (2.00,2.80) rectangle (3.00,3.15); \draw[black,thin] (2.00,2.80) rectangle (3.00,3.15); \node[font=\footnotesize,black] at (2.50,2.97) {4.11};
        \fill[colorunstable] (3.00,2.80) rectangle (4.00,3.15); \draw[black,thin] (3.00,2.80) rectangle (4.00,3.15); \node[font=\footnotesize,black] at (3.50,2.97) {4.23};
        \fill[colorunstable] (4.00,2.80) rectangle (5.00,3.15); \draw[black,thin] (4.00,2.80) rectangle (5.00,3.15); \node[font=\footnotesize,black] at (4.50,2.97) {4.21};
        \fill[colorunstable] (5.00,2.80) rectangle (6.00,3.15); \draw[black,thin] (5.00,2.80) rectangle (6.00,3.15); \node[font=\footnotesize,black] at (5.50,2.97) {4.12};

        \fill[colorstable] (0.00,3.15) rectangle (1.00,3.50); \draw[black,thin] (0.00,3.15) rectangle (1.00,3.50); \node[font=\footnotesize,black] at (0.50,3.32) {4.75};
        \fill[colorstable] (1.00,3.15) rectangle (2.00,3.50); \draw[black,thin] (1.00,3.15) rectangle (2.00,3.50); \node[font=\footnotesize,black] at (1.50,3.32) {4.98};
        \fill[colorunstable] (2.00,3.15) rectangle (3.00,3.50); \draw[black,thin] (2.00,3.15) rectangle (3.00,3.50); \node[font=\footnotesize,black] at (2.50,3.32) {5.08};
        \fill[colorunstable] (3.00,3.15) rectangle (4.00,3.50); \draw[black,thin] (3.00,3.15) rectangle (4.00,3.50); \node[font=\footnotesize,black] at (3.50,3.32) {5.21};
        \fill[colorunstable] (4.00,3.15) rectangle (5.00,3.50); \draw[black,thin] (4.00,3.15) rectangle (5.00,3.50); \node[font=\footnotesize,black] at (4.50,3.32) {5.22};
        \fill[colorunstable] (5.00,3.15) rectangle (6.00,3.50); \draw[black,thin] (5.00,3.15) rectangle (6.00,3.50); \node[font=\footnotesize,black] at (5.50,3.32) {5.47};

        \fill[colorstable] (0.00,3.50) rectangle (1.00,3.85); \draw[black,thin] (0.00,3.50) rectangle (1.00,3.85); \node[font=\footnotesize,black] at (0.50,3.67) {5.72};
        \fill[colorstable] (1.00,3.50) rectangle (2.00,3.85); \draw[black,thin] (1.00,3.50) rectangle (2.00,3.85); \node[font=\footnotesize,black] at (1.50,3.67) {5.93};
        \fill[colorstable] (2.00,3.50) rectangle (3.00,3.85); \draw[black,thin] (2.00,3.50) rectangle (3.00,3.85); \node[font=\footnotesize,black] at (2.50,3.67) {6.14};
        \fill[colorstable] (3.00,3.50) rectangle (4.00,3.85); \draw[black,thin] (3.00,3.50) rectangle (4.00,3.85); \node[font=\footnotesize,black] at (3.50,3.67) {6.33};
        \fill[colorstable] (4.00,3.50) rectangle (5.00,3.85); \draw[black,thin] (4.00,3.50) rectangle (5.00,3.85); \node[font=\footnotesize,black] at (4.50,3.67) {6.53};
        \fill[colorunstable] (5.00,3.50) rectangle (6.00,3.85); \draw[black,thin] (5.00,3.50) rectangle (6.00,3.85); \node[font=\footnotesize,black] at (5.50,3.67) {6.53};

        \node[below,font=\small] at (0.50,0) {18};
        \node[below,font=\small] at (1.50,0) {20};
        \node[below,font=\small] at (2.50,0) {22};
        \node[below,font=\small] at (3.50,0) {24};
        \node[below,font=\small] at (4.50,0) {26};
        \node[below,font=\small] at (5.50,0) {28};
        \node[below,font=\normalsize] at (3.00,-0.7) {$n_{\mathrm{initial}}$};

        \end{tikzpicture}
        \hspace*{0.4cm}(\textit{b}) Angular momentum flux $J^{\omega}/J^{\omega}_{0}$
    \end{minipage}
    \caption{Roll stability map for $400 \leq \Rey \leq 3 \times 10^3$ in the setup of 
    \cite{martinez-ariasEffectNumberVortices2014} for $\Gamma=30$ and $\eta=0.909$. 
    (\textit{a}) Stable ({\color{colorstable}\rule{0.3cm}{0.2cm}}) versus unstable ({\color{colorunstable}\rule{0.3cm}{0.2cm}}) 
    configurations with initial and final number of rolls. 
    (\textit{b}) Corresponding normalized angular momentum flux $J^{\omega}/J^{\omega}_{0}$ for each configuration.
    Note that after surpassing the chaotic wavy vortex flow regime, more rolls regain their stability.} 
    \label{fig:roll_stability_maps}
\end{figure}

\noindent Figure \ref{fig:martinez} illustrates the occurrence of multiple states across the investigated Reynolds number range.
The presented phase space exhibits distinct regimes with qualitatively different behavior. 
At low Reynolds numbers $(\Rey < 900)$, multiple stable states coexist, and the final state
depends on initialization, indicating a hysteresis region with significant accessible phase
space volume. In the lower intermediate regime $(900 < \Rey < 1250)$, the system converges
to a single state, but with fewer rolls than at low Re, suggesting roll merging events. The
upper intermediate regime $(1250 < \Rey < 2000)$ exhibits unique final states regardless of
initial conditions, representing a region of strong convergence.
At high Reynolds numbers $(\Rey > 2500)$, we observe a remarkable regrowth of the accessible phase space region for
possible roll configurations.
This phenomenon is of particular interest. As described in the work of
\cite{dutcherSpatiotemporalModeDynamics2009}, the expansion of accessible phase
space at high $\Rey$ highlights the crucial role of flow dynamics in determining the stability of
respective flow patterns. The system undergoes a cascade of transitions from TVF through WVF, modulated wavy vortex 
flow (MWF), chaotic wavy vortex flow (cWVF), turbulent wavy vortex flow (tWVF), to turbulent Taylor--vortex flow (tTVF). 
This cascade provides insight into the regrowth dynamics. At sufficiently high
Reynolds numbers, turbulence might suppress or dissolve secondary instabilities, therefore restabilizing large-scale 
coherent structures. Additionally, 
no-slip boundary conditions preferentially stabilize roll structures near the endwalls, including another persistent 
stabilization effect across several $\Rey$-regimes even in otherwise turbulent flow 
(see \cite{rameshSuspensionTaylorCouette2019} and section \ref{sec_flow_regimes} for detailed flow structure analysis).

The stability map (figure \ref{fig:roll_stability_maps}) clearly delineates persistent (magenta) and 
non-persistent (cyan) regions in $(\Rey, n_{\text{initial}})$ space. The embedded numbers show the 
actual roll numbers achieved after the system reaches statistical steady state, revealing the complex 
structure of the attractor landscape.
Persistent configurations (appearing as magenta regions in figure \ref{fig:roll_stability_maps} \textit{a}), 
indicate that certain $(\Rey, n_\mathrm{initial})$ 
combinations are absolutely stable and resist perturbations. Reduction channels manifest
as cyan regions showing systematic roll number reduction through merging events. Furthermore, critical
boundaries separate persistent and non-persistent configurations with sharp transitions, suggesting
that the system exhibits bifurcation-like behavior at these interfaces.

\subsection{Roll aspect ratio effect on transport}

\begin{figure}[h!]
    \centering
    \includegraphics{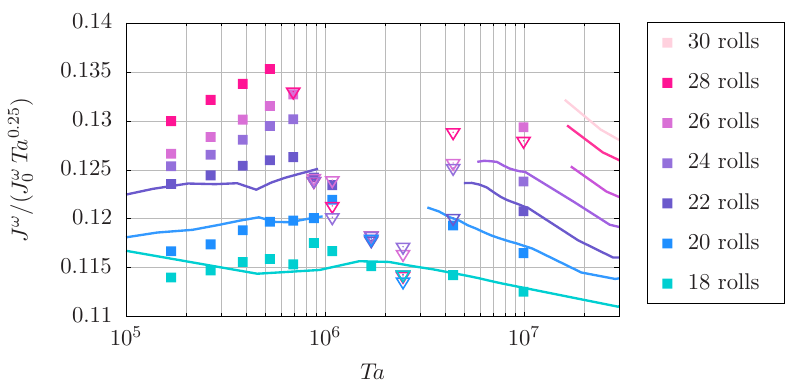}
    \caption{Angular momentum flux development for $10^5 \leq T\!a < 3 \times 10^7$ in the 
    setup with $\Gamma=30$ and $\eta=0.909$ normalized by $T\!a^{0.25}$. 
    Solid lines denote the experimental data by \cite{martinez-ariasEffectNumberVortices2014} for 
    different stable roll configurations. Coloured squares (triangles) show our DNS results 
    of multiple persistent (non-persistent) states. Colours were chosen as in figure \ref{fig:martinez}.}
    \label{fig:martinez_normalized}
\end{figure}

Following \cite{martinez-ariasEffectNumberVortices2014}, we investigated the scaling of angular
momentum flux with Taylor number for different roll configurations. Our results 
(see figures \ref{fig:martinez} and \ref{fig:martinez_normalized}) highlight that the angular momentum flux efficiency increases
with the number of rolls in the investigated Taylor number range, consistent with theoretical expectations. 
This behaviour differs from that observed in the ultimate regime \citep{martinez-ariasEffectNumberVortices2014}.

Our analysis also demonstrates that different roll states exhibit distinct scaling exponents
 $\alpha$ in the relationship $J^\omega \propto \mathit{Ta}^\alpha$. 
 
\noindent This finding is consistent with previous studies that identified the dependence of
the proportionality coefficient on the aspect ratio of individual rolls \citep{froitzheimAngularMomentumTransport2019}.
Interestingly, this behavior differs from that observed in DNS of two-dimensional Rayleigh--
Bénard convection, where the scaling exponent remained invariant across different roll
configurations \citep{wangMultipleStatesTurbulent2020}. Beyond the inherent differences between 3D and 2D flow structures, 
the difference may arise from the additional geometric constraints imposed by
the cylindrical geometry and the no-slip endwalls in Taylor--Couette systems, which break
the strict two-dimensional symmetry present in idealized 2D Rayleigh--Bénard configurations with periodic side walls.
This systematic variation of the transport scaling with roll configuration
provides direct evidence that the aspect ratio of coherent structures, and not only the global
Reynolds number, plays a fundamental role in determining turbulent transport efficiency in Taylor--Couette flow.

\subsection{Transition dynamics and energy budget analysis}
\label{sec_flow_regimes}
To gain deeper insight into the mechanisms governing transitions between different roll
configurations, we analyze the temporal evolution of unstable states and quantify the energy
transfer between dominant flow modes.

\subsubsection{Space-time dynamics of unstable configurations}

\begin{figure}[h!]
    \centering
    \begin{minipage}[c]{0.4in}
        \centering
        \hspace*{-7.0cm}
        \begin{tikzpicture}
            \node[anchor=south west,inner sep=0] (image) at (0,0) 
                {\includegraphics{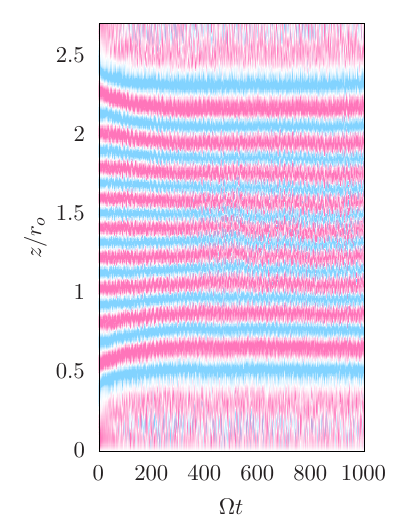}};
            \node[anchor=north west, yshift=5pt, xshift=1.1cm] at (image.north west) {(\textit{a})};
        \end{tikzpicture}
    \end{minipage}
    \hspace{-0.5cm}
    \begin{minipage}[c]{0.4in}
        \centering
        \begin{tikzpicture}
            \node[anchor=south west,inner sep=0] (image) at (0,0) 
                {\includegraphics{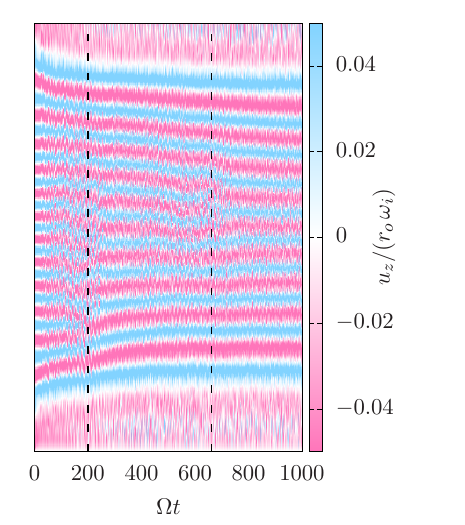}};
            \node[anchor=north west, yshift=5pt] at (image.north west) {(\textit{b})};
        \end{tikzpicture}
    \end{minipage}
    \caption{Space--time diagrams for two different initial roll developments in the setup of 
    \cite{martinez-ariasEffectNumberVortices2014} for $\Gamma=30$ and $\eta=0.909$. 
    Shown is the normalized axial velocity $u_z$ at mid-gap radius and $\phi = 0.$\\
    (a) Persistent configuration for $\Rey = 1250$ and $n_{\text{initial}}=20$. 
    (b) Non-persistent initialisation $\Rey = 1250$, $n_{\text{initial}}=26$ with pronounced roll merging 
    occurrences highlighted by the dotted lines.} 
    \label{fig:spacetime}
\end{figure}

Figure \ref{fig:spacetime} shows space-time diagrams of the axial velocity 
component $u_z(r,\phi,z,t)$ at mid-gap radius and $\phi=0$ for two different configurations. These diagrams 
provide valuable insight into the characteristic
evolution of roll merging events and allow for systematic tracking of the temporal dynamics. 

\newpage
In contrast to persistent configurations, which exhibit
stationary horizontal bands with consistent spacing (figure \ref{fig:spacetime}\textit{a}), 
non-persistent states (figure \ref{fig:spacetime}\textit{b}) display pronounced diagonal 
features indicating temporal evolution of the roll structures and their progressive reorganization.

We observe systematic merging events occurring over extended timescales of $\Delta t \approx 200$ to $400$ rotation periods, 
demonstrating the relatively slow nature of these instability-driven transitions. 
The merging process proceeds through well-defined stages that can be clearly identified in the space-time diagrams. 
Two neighboring rolls, rotating in the same direction,
gradually approach each other axially over time, the interface between them progressively weakens and becomes 
increasingly diffuse until the
two rolls coalesce into a single, larger vortex structure with reduced angular momentum transport. This process 
continues iteratively until the
system reaches a stable configuration with fewer rolls and increased axial wavelength.

The spatial location of merging events is not random but shows clear preferential occurrence
away from the endwalls, where no-slip boundary conditions and geometric constraints stabilize present roll flow 
structures and suppress local instabilities. 
This observation aligns with our earlier discussion (section \ref{sec:validation}) 
regarding the crucial role of realistic boundary conditions in determining flow evolution and the importance 
of resolving endwall effects accurately. 

\subsubsection{Modal energy transfer during roll merging}
To quantify the energy redistribution during transitions, we perform azimuthal Fourier
decomposition of the velocity field. 
The analysis is restricted to the radial velocity component, which exhibits the strongest signature of the 
vortex roll dynamics and azimuthal waviness,

\begin{equation}
    u_r(r, \phi, z, t) = \sum_{m=0}^{M} \hat{u}_{r,m}(r, z, t) e^{im\phi},
\end{equation}

\noindent where $m$ is the azimuthal wavenumber and $M$ is the maximum wavenumber considered in the truncated Fourier series. 

We further compute the radial fraction of the kinetic energy associated with each mode:

\begin{equation}
    E_{r,m}(t) = \frac{1}{2H}\int_z |\hat{u}_{r,m}|^2 \bigg|_{r=\frac{r_i+r_o}{2}, \phi=0} \mathrm{d}z.
\end{equation}

\begin{figure}[H]
    \centering

    \begin{tikzpicture}
        \node[anchor=south west,inner sep=0] (image) at (0,0) 
            {\includegraphics{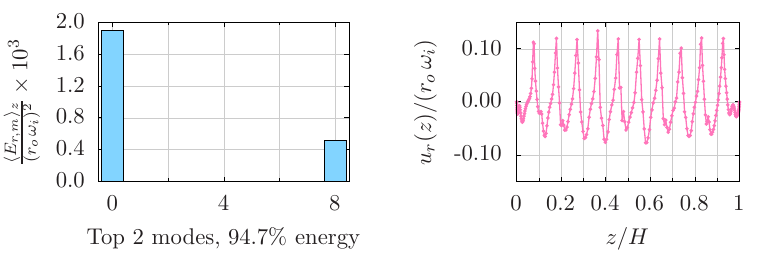}};
        \node[anchor=north west, yshift=5pt] at (image.north west) {(\textit{a})};
    \end{tikzpicture}
    
    \begin{tikzpicture}
        \node[anchor=south west,inner sep=0] (image) at (0,0) 
            {\includegraphics{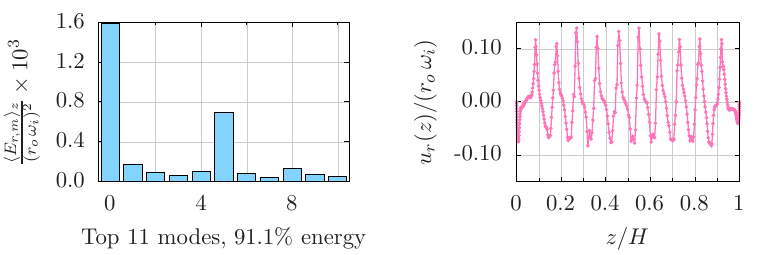}};
        \node[anchor=north west, yshift=5pt] at (image.north west) {(\textit{b})};
    \end{tikzpicture}
    
    \begin{tikzpicture}
        \node[anchor=south west,inner sep=0] (image) at (0,0) 
            {\includegraphics{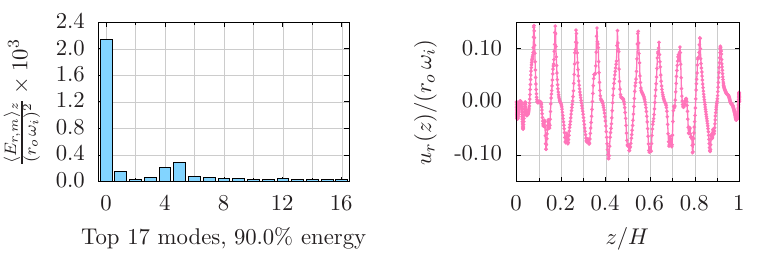}};
        \node[anchor=north west, yshift=5pt] at (image.north west) {(\textit{c})};
    \end{tikzpicture}
    
    \begin{tikzpicture}
        \node[anchor=south west,inner sep=0] (image) at (0,0) 
            {\includegraphics{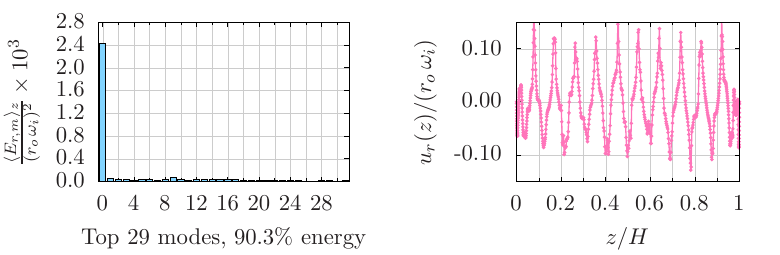}};
        \node[anchor=north west, yshift=5pt] at (image.north west) {(\textit{d})};
    \end{tikzpicture}
    
    \caption{Mode decompositions of the radial kinetic energy fraction in the 
    setup of \cite{martinez-ariasEffectNumberVortices2014} for $\Gamma=30$, 
    $\eta=0.909$ and $n=20$, as obtained in the DNS for 
    (\textit{a}) $\Rey = 800$
    (\textit{b}) $\Rey = 1250$
    (\textit{c}) $\Rey = 2000$
    (\textit{d}) $\Rey = 3000$. Axial profiles of the normalized radial velocities for the respective cases are shown on the right.} 
    \label{fig:modenamplitude}
\end{figure}
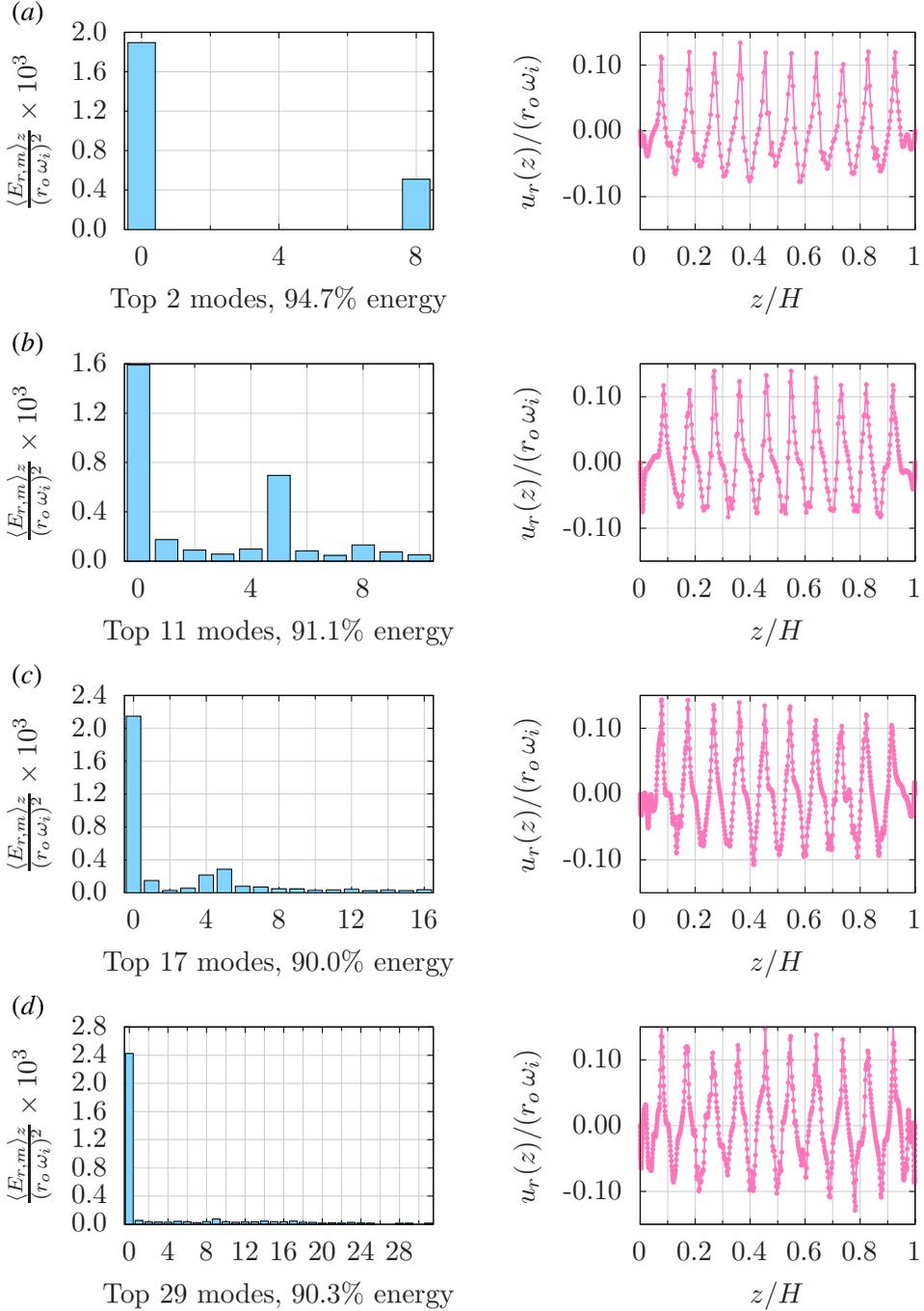

\noindent Figure \ref{fig:modenamplitude} shows the energy distribution among 
azimuthal modes for four persistent roll configurations at different $\Rey$, together 
with the corresponding axial profiles of $u_r$. Table \ref{tab:mode_energy} 
presents the energy fractions carried by the axisymmetric mode ($m=0$) and 
non-axisymmetric modes ($m > 0$), respectively. At $\Rey=800$, the flow exhibits weak wavy vortex flow (WVF), with the 
axisymmetric mode dominant and only one additional azimuthal mode weakly excited. 
At $\Rey=1250$, the azimuthal modulation intensifies significantly, with 
non-axisymmetric modes now carrying more energy than the previously dominant 
$m=0$ mode. At $\Rey=2000$, a resurgence of the axisymmetric mode is observed, 
with $m=0$ regaining dominance. This re-emergence of axisymmetric structure 
in the turbulent regime is consistent with the observations of 
\citet{dutcherSpatiotemporalModeDynamics2009}, who described a cascade of 
transitions from WVF through modulated wavy vortex flow (mWVF) and chaotic 
wavy vortex flow (cWVF) towards turbulent wavy vortex flow (tWVF). 
\newpage 

\newpage
\begin{longtable}{r@{\hspace{1cm}}c@{\hspace{1cm}}c}
\label{tab:mode_energy}\\
$\Rey$ & $m=0$ (\%) & $m> 0$ (\%) \\
\hline
\\[-3mm]  
\endfirsthead

\multicolumn{3}{r}{{\tablename\ \thetable{} -- continued}}\\
\hline
$\Rey$ & $m=0$ (\%) & $m> 0$ (\%) \\
\hline
\\[-8mm]  
\endhead

\hline
\endfoot

\endlastfoot
800  & 74.5 & 25.5 \\
1250 & 46.9 & 53.1 \\
2000 & 67.2 & 32.8 \\
3000 & 60.5 & 39.5 \\
\hline
\caption{Energy fractions of zeroth- and higher-order modes}
\end{longtable}

\vspace{-0.2cm}
\noindent Our findings support this interpretation: the corresponding phase space analysis 
reveals that the stability region contracts and expands in correlation with these 
transitions, suggesting that variations in azimuthal waviness directly influence 
the extent of the persistent parameter space.

\begin{figure}[h!]
    \centering
    \includegraphics{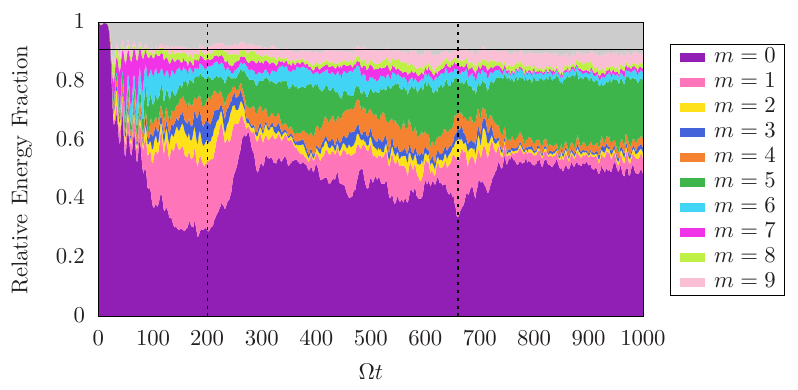}
    \caption{Energy distribution development for $\Rey = 1250$, $n_{\text{init}}=26$ in the 
    setup of \cite{martinez-ariasEffectNumberVortices2014}.
    The top 10 modes, which together cover over 90\% of the system's total energy budget, are 
    shown with distinct colours. In this non-persistent configuration, several energy shifts
    occur and subsequently push the system into another configuration scheme. Dotted lines 
    highlight completed roll merging events as shown in figure \ref{fig:spacetime}.}
    \label{fig:energy_distribution_re2000}
\end{figure}

Figure \ref{fig:energy_distribution_re2000} displays the temporal evolution of modal energy fractions
for a representative transition case ($\Rey = 1250$, $n_{\text{initial}} = 26 \to n_{\text{final}} = 20$). 
Several key features emerge from this analysis. The axisymmetric mode ($m=0$) remains dominant throughout the transition, 
which reflects the persistence of 
the basic vortex structures even during reorganization. However, non-axisymmetric modes ($m > 0$) exhibit transient 
amplification during merging events, emphasized by the dotted lines in figures \ref{fig:spacetime} 
and \ref{fig:energy_distribution_re2000}, while the dominant non-axisymmetric mode shifts during the transition process.
This highlights that different roll configurations in the given setup might also lead to a different 
distribution of excited frequencies in the regard of secondary instabilities. 
Finally, after the transition completes, the modal energy distribution reaches a new equilibrium
corresponding to the final roll configuration. Interestingly, each transition event is preceded by a 
reduction in the $0$-mode's energy fraction, indicating that the system draws energy from the base 
flow to realize configurational changes. 
This phenomenon could serve as a predictor for transition scenarios and will be part of future work.

\section{Conclusions}
\label{sec:conclusion}
We have performed a DNS study of Taylor--Couette flow with realistic no-slip boundary conditions at all surfaces. 
In this study, we extended the classical periodic angular momentum flux formulation by \cite{eckhardtTorqueScalingTurbulent2007} 
to include axial transport terms arising from $z$-dependent azimuthal velocity. This modification yields spatially constant 
angular momentum fluxes and improved boundary layer predictions. \\
The wall-lid singularity inherent in this configuration requires careful treatment via smoothing functions. 
We demonstrated that sigmoid smoothing provides numerical stability while introducing only a constant offset 
in angular momentum flux that can be systematically corrected.
Excellent quantitative agreement with experimental data from \cite{martinez-ariasEffectNumberVortices2014} 
and \cite{rameshSuspensionTaylorCouette2019} validates our DNS approach and confirms the experimental 
relevance of the simulated flow structures.

A systematic parameter space exploration revealed coexistence of multiple stable flow states with 
different roll numbers at identical Reynolds numbers. Stability diagrams provide a foundation for future investigations into the 
detailed mechanisms governing state selection and transitions.
The system exhibits strong hysteresis, with the final state depending on initial conditions. This 
multistability arises from competing length scales, Ekman layer effects, and finite-size disturbances, 
wherein different roll configurations interact distinctly with endwall boundary layers, creating separate attractors in phase space.

The existence of multiple stable states with different transport efficiencies has important implications 
for both fundamental understanding and practical applications. From a fundamental perspective, it demonstrates 
that Taylor--Couette flow, even with fully deterministic governing equations and fixed control parameters, does 
not possess a unique turbulent state. Rather, the system exhibits history-dependent behavior, with the initial 
conditions determining which attractor is realized.
From a practical perspective, this multistability presents opportunities for flow control. Preferential selection 
of states with higher transport efficiency (typically those with more smaller rolls) could optimize industrial 
mixing processes or heat transfer applications. Conversely, understanding which perturbations trigger transitions 
between states could help prevent undesired regime changes in sensitive applications.
Future work will focus on identifying the physical mechanisms governing transitions between multiple states in 
Taylor--Couette flow and their relevance to other complex hydrodynamic systems.

\section*{Acknowledgments}
We truly appreciate the many fruitful discussions with D. Lohse, G. Vacca and H. Dave during the course of this work. 
Experimental validations were made possible by the support of M. Alam, J. Peixinho and P. Ramesh, 
who kindly shared their measurement data with us.
We further acknowledge financial support from German Research Foundation (DFG) under grants Sh405/20, Sh405/22 and SO2399/2.
The presented numerical investigations were enabled by the supercomputing clusters of the Max Planck Computing and Data Facilities.

\newpage

\section*{Appendix}

\renewcommand{\thesubsection}{A.\arabic{subsection}}
\subsection{Flow field initialization}
\label{seq:appendix}

To initialize DNS with a Taylor vortex flow that satisfies no-slip boundary conditions at the endplates, and the 
continuity equation is not trivial.
Therefore, we prioritize the exact enforcement of no-slip boundary conditions at the endplates, accepting a 
minor deviation from perfect divergence-free conditions of the initial flow field, which is localized to the 
narrow smoothing regions (2\% of the domain height at each boundary).
Thus the initial field consists of a base Couette flow with superimposed Taylor vortex perturbations.
The velocity components in cylindrical coordinates $(r, \varphi, z)$ are given by:

\begin{align}
u_\varphi(r,z) &= S_z(z) \left[ \frac{\eta^2}{\eta^2-1}\left(r - \frac{1}{r}\right) - \frac{A}{2}F(r)\sin(\Phi) \right], \label{eq:uphi_init} \\
u_r(r,z) &=  \frac{A}{r}n\pi  F(r)S_z(z) \frac{\mathrm{d}z'}{\mathrm{d}z} \sin(\Phi), \label{eq:ur_init} \\
u_z(r,z) &=  \frac{A}{r} \frac{\mathrm{d}F}{\mathrm{d}r}S_z(z) \cos(\Phi), \label{eq:uz_init}
\end{align}

\noindent where $\eta = r_{i}/r_{o}$, $A = 0.05\eta^2$ is the perturbation amplitude, and $n$ is the number of imposed rolls. 
The phase function is defined as

\begin{equation}
\Phi(z) = n\pi z'(z) - \frac{\pi}{2},
\end{equation}

\noindent with a simple axial coordinate transformation $z'(z)$ described below.
The radial shape function $F(r)$ ensures no-slip conditions at both cylinder walls.
We employ

\begin{equation}
F(r') = {r'}^2(1-r')^2, \quad \text{with} \quad r' = \frac{r-\eta}{1-\eta},
\label{eq:shape_function}
\end{equation}

\noindent whose derivative is

\begin{equation}
\frac{\mathrm{d}F}{\mathrm{d}r} = \frac{2r'(1-r')(1-2r')}{1-\eta}.
\end{equation}

\noindent This choice guarantees that both $F$ and $\mathrm{d}F/\mathrm{d}r$ vanish at $r'=0$ (inner cylinder) and $r'=1$ (outer cylinder), 
thereby satisfying $u_r = u_z = 0$ at both walls.
To impose no-slip conditions at the endplates ($z=0$ and $z=H$), we introduce a sigmoid-based smoothing 
function with min-max-normalization

\begin{equation}
S_z(z) = \frac{M_{\text{down}}(z) \cdot M_{\text{up}}(z) - \min}{\max - \min},
\label{eq:sigmoid}
\end{equation}

\noindent where

\begin{align}
M_{\text{down}}(z) &= \frac{1}{1 + \exp[(z - z_{0,\text{down}})/\epsilon]}, \\
M_{\text{up}}(z) &= \frac{1}{1 + \exp[-(z - z_{0,\text{up}})/\epsilon]}.
\end{align}

\noindent Here, $\epsilon = 0.02H$ defines the boundary layer thickness (2\% of the domain height), 
and $z_{0,\text{down}} = 0.01H$ and $z_{0,\text{up}} = 0.99H$ are the transition centers. 
The terms $\min$ and $\max$ denote the minimum and maximum of $M_{\text{down}}(z) M_{\text{up}}(z)$ over $z \in [0,H]$, 
which rescales the smoothing function to $S_z(z) \in [0,1]$.
After normalization, we explicitly set $S_z(0) = S_z(H) = 0$, ensuring that all velocity components vanish at the endplates.
For uniformly sized vortex rolls along the axial direction, we employ a simple linear mapping

\begin{equation}
z'(z) = \frac{z}{H},
\label{eq:z_mapping}
\end{equation}

\noindent which distributes all $n$ vortex rolls equally over the domain height $H$.\\
The constructed velocity field satisfies the following boundary conditions:
\begin{alignat}{4}
    &\text{Inner cylinder} \,(r=r_1)\!: \hspace{0.5cm}& u_r &= 0,\hspace{0.25cm} & u_\phi &= S_z(z) \eta,\hspace{0.25cm} & u_z &= 0 \quad \text{at } r = r_1, \notag \\
    &\text{Outer cylinder} \,(r=r_2)\!: \hspace{0.5cm}& u_r &= 0,\hspace{0.25cm} & u_\phi &= 0,\hspace{0.25cm} & u_z &= 0 \quad \text{at } r = r_2, \notag \\
    &\text{Top, bottom lids} \,(z=0, z=H)\!: \hspace{0.5cm} & u_r &= 0,\hspace{0.25cm} & u_\phi &= 0,\hspace{0.25cm} & u_z &= 0 \quad \text{at } z = 0, H.
\end{alignat}

\noindent The azimuthal velocity at the inner cylinder reproduces the exact Couette profile

\begin{equation}
u_\varphi(r=\eta, z) = S_z(z) \eta = S_z(z) \omega_1 r_1,
\end{equation}

\noindent modulated by the axial smoothing function to satisfy the no-slip condition at the endplates.
The radial and axial velocity components are derived from a stream function $\psi(r,z) = A F(r) S_z(z) \cos(\Phi)$ via

\begin{equation}
u_r = \frac{1}{r}\frac{\partial \psi}{\partial z}, \quad u_z = -\frac{1}{r}\frac{\partial \psi}{\partial r},
\end{equation}

\noindent which, under the approximation that $\frac{\mathrm{d} S(z)}{\mathrm{d} z}$ is neglicible for the majority of the domain, automatically satisfies the continuity equation

\begin{equation}
\nabla \cdot \mathbf{u} = \frac{1}{r}\frac{\partial (r u_r)}{\partial r} + \frac{\partial u_z}{\partial z} = 0.
\end{equation}

\noindent To trigger the transition to wavy vortex flow or turbulent states, 
we add white noise to the initial velocity field. To ensure that the perturbed field remains almost divergence-free, we employ a stream function approach for the radial and axial velocity components, while the azimuthal component is perturbed independently.
We introduce a random stream function $\vartheta$ and its spatial derivatives at each grid point $(j,k,i)$:

\begin{align}
\vartheta_{ki} &\propto \mathcal{U}(-\zeta \eta, \zeta \eta), \\
\left(\frac{\partial \vartheta}{\partial r}\right)_{ki} &\propto \mathcal{U}(-\zeta \eta, \zeta \eta), \\
\left(\frac{\partial \vartheta}{\partial z}\right)_{ki} &\propto \mathcal{U}(-\zeta \eta, \zeta \eta),
\end{align}

\noindent where $\zeta = 0.01$ is the noise intensity (1\% of the characteristic velocity $\eta$ in the non-dimensionalized setting), and $\mathcal{U}(a,b)$ denotes a uniform distribution on the interval $[a,b]$. The noise is then applied to the velocity components as:

\begin{align}
u_r^\text{(noisy)}(j,k,i) &= u_r(j,k,i) - \frac{1}{r} \frac{\partial \vartheta}{\partial z}, \\
u_z^\text{(noisy)}(j,k,i) &= u_z(j,k,i) + \frac{1}{r} \frac{\partial \vartheta}{\partial r}, \\
u_\varphi^\text{(noisy)}(j,k,i) &= u_\varphi(j,k,i) + \xi_{jki},
\end{align}

\noindent where $\xi_{jki} \propto \mathcal{U}(-\zeta \eta, \zeta \eta)$ is an independent random perturbation for the azimuthal component. This stream function formulation automatically satisfies the incompressibility constraint $\nabla \cdot \mathbf{u} = 0$ for the radial and axial components in cylindrical coordinates, as

\begin{equation}
\frac{1}{r}\frac{\partial (r u_r)}{\partial r} + \frac{\partial u_z}{\partial z} 
= \frac{1}{r}\frac{\partial}{\partial r}\left(-\frac{\partial \vartheta}{\partial z}\right) + \frac{\partial}{\partial z}\left(\frac{1}{r}\frac{\partial \vartheta}{\partial r}\right) = 0.
\end{equation}

\noindent The random numbers are generated independently at each grid point, providing spatially uncorrelated perturbations across all directions.
This initialization procedure yields an almost divergence-free velocity field that satisfies all physical boundary conditions while containing the characteristic Taylor vortex structure with realistic boundary layers near the endplates.

\subsection{Angular momentum flux: extension for no-slip boundaries}
\label{seq:appendix_angular_flux}
\begin{figure}[h!]
    \centering
    \includegraphics{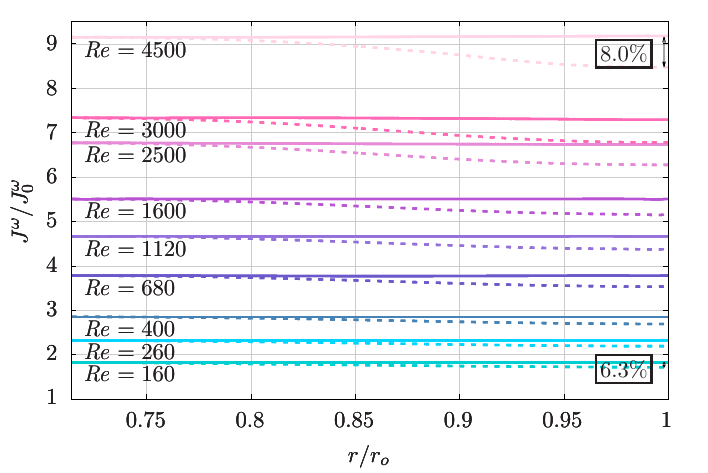}
    \caption{Radial dependence of normalised angular momentum flux for $160 \leq \Rey \leq 4.5 \times 10^3$ 
    in the 
    setup of \cite{ostillaOptimalTaylorCouette2013} for $\Gamma=2\pi$ and $\eta=5/7$.  
    Differences between the periodic approximation and the full no-slip consideration approaches are pronounced. 
    Contribution of the ``periodic'' past , i.e. $\langle J_r\rangle_{z,t}$ in total $J^\omega$ is dominant, 
    but not sufficient for accurate calculation of $J^\omega$, which
    must be constant for all values of $r.$}
    \label{fig:angular_momentum_flux}
\end{figure}

We now validate the necessity and correctness of the introduced and extended angular momentum flux formulation.
Figure \ref{fig:angular_momentum_flux} compares the radial profiles of $J^\omega(r)/J_0^\omega$ computed using two different methods. 
The first approach employs the periodic boundary approximation (figure \ref{fig:angular_momentum_flux} dotted lines), which neglects axial 
transport terms. This formulation yields:
\begin{equation}
    J_{\text{periodic}}^\omega = r^3\left(\langle u_r\omega\rangle_{A,t}-\nu\partial_r\langle\omega\rangle_{A,t} \right).
\end{equation}

\noindent The second approach uses the no-slip boundary formulation (figure \ref{fig:angular_momentum_flux} solid lines), which 
includes the axial correction term $J_{\text{axial}}$ derived in section \ref{sec:system}, accounting for 
z-dependent angular momentum transport through Ekman layers:
\begin{equation}
    J_{\text{total}}^\omega = r^3\left(\langle u_r\omega\rangle_{A,t}-\nu\partial_r\langle\omega\rangle_{A,t} \right) +\int_{r_1}^r r'^3\left(\langle\partial_z(u_z\omega)\rangle_{A,t}-\nu\langle\partial_z^2(\omega)\rangle_{A,t}\right)\mathrm{d}r'.
\end{equation}
\noindent The periodic approximation produces angular momentum flux profiles that 
vary strongly with radius. This radial dependence is physically
inconsistent, since a statistically steady state with no internal sources or sinks demands a conserved 
and therefore constant angular momentum flux across the gap.
In contrast, the extended formulation including axial terms yields
constant angular momentum flux profiles across the gap for all Reynolds numbers tested. 
This behavior is maintained throughout the parameter range $\Rey \in [160, 4500]$, including multiple flow regimes from 
laminar Taylor vortex flow to turbulent Taylor--vortex flow.

The necessity of the axial correction term $J_{\text{axial}}$ demonstrates that Ekman layer pumping at the endwalls 
significantly redistributes angular momentum in the axial direction. Neglecting this redistribution violates 
conservation and leads to anomalous radial gradients in computed transport quantities. Our results confirm that 
proper treatment of no-slip boundary conditions requires explicit accounting for three-dimensional momentum transport, 
beyond what is captured by periodic approximations.

\newpage

\subsection{Resolution study}
\label{seq:appendix_resolution}

\begin{figure}[h!]
    \vspace{-0.2cm}
    \centering
    \includegraphics{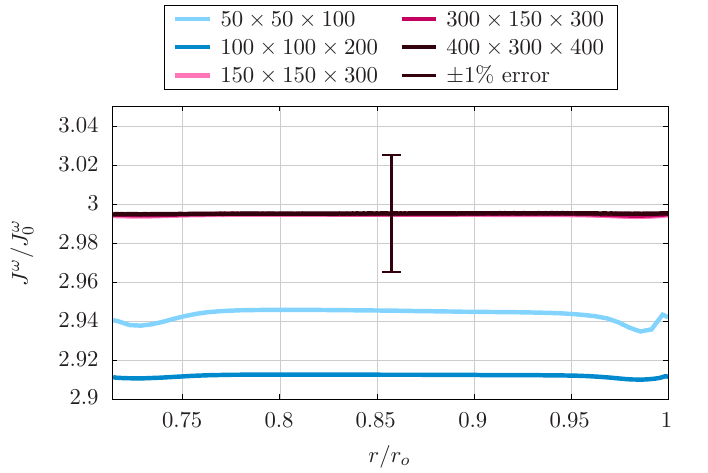}
    \caption{Radial dependence of normalised angular momentum flux for five different grid 
    resolutions ($N_\phi\times N_r\times N_z$). 
    Control parameters were chosen as in Ostilla \textit{et al.} (2013): $\mathit{Ta} = 2.44 \times 10^5$, $\Gamma=2\pi$, $\eta=5/7$, 
    with no-slip endwall boundary conditions. A 1\% error bar is presented, 
    highlighting the sufficiently well resolved cases.}
    \label{fig:grid_convergence}
\end{figure}
\newpage

\noindent Figure \ref{fig:grid_convergence} shows results of a grid resolution 
study for $\mathit{Ta} = 2.44\times 10^5$, corresponding to the setup studied by Ostilla \textit{et al.} (2013), 
but with no-slip endwalls. The results reveal that the two lowest resolution configurations deliver inacurrate angular momentum 
flux measurements, which directly highlights the discrepancies that can occur 
when the smoothing region is insufficiently resolved.

\subsection{Simulation results}

\setlength\LTleft{0pt plus 1fill}
\setlength\LTright{0pt plus 1fill}
\small
\noindent 
\begin{longtable}{c c c c c c c c c}
\caption{Overview of conducted direct numerical simulations. Listed are: Reynolds number $\Rey$, 
Taylor number $\mathit{Ta}$, grid resolution $N_\phi \times N_r \times N_z$, initial number 
of rolls $n_{\text{initial}}$, aspect ratio $\Gamma$, normalized angular momentum flux $J^{\omega}/J^{\omega}_{0}$, 
total simulation time $t_\text{total}$ in rotational time units, maximum grid spacing to Kolmogorov length ratio 
$\max\langle \updelta/\eta\rangle$ (averaged over $\phi,z, t$), and corresponding section in text.}
\\

$\Rey$ & $\mathit{Ta}$ & $N_\phi \times N_r \times N_z$ & $n_{\text{initial}}$ & $\Gamma$ &
$J^{\omega}/J^{\omega}_{0}$ & $t_{\text{total}}$ &
$\max\langle \updelta/\eta\rangle_{\phi,z,t}$ & Section \\
\hline
\endfirsthead

$\Rey$ & $\mathit{Ta}$ & $N_\phi \times N_r \times N_z$ & $n_{\text{initial}}$ & $\Gamma$ &
$J^{\omega}/J^{\omega}_{0}$ & $t_{\text{total}}$ &
$\max\langle \updelta/\eta\rangle_{\phi,z,t}$ & Section \\
\hline
\endhead

\hline
\multicolumn{9}{r}{Continued on next page} \\
\endfoot

\hline
\endlastfoot
$100$ & $1.108 \times 10^{4}$ & $256 \times 64 \times 256$ & $0$ & $30$ & $0.949$ & $900$ & $0.185$ &  $4.1$ \\
$200$ & $4.431 \times 10^{4}$ & $256 \times 64 \times 256$ & $0$ & $30$ & $1.615$ & $900$ & $0.268$ &  $4.1$ \\
$300$ & $9.969 \times 10^{4}$ & $256 \times 64 \times 256$ & $0$ & $30$ & $2.189$ & $900$ & $0.356$ &  $4.1$ \\
$400$ & $1.772 \times 10^{5}$ & $256 \times 64 \times 256$ & $0$ & $30$ & $2.599$ & $900$ & $0.442$ &  $4.1$ \\
$500$ & $2.769 \times 10^{5}$ & $256 \times 64 \times 256$ & $0$ & $30$ & $2.952$ & $900$ & $0.531$ &  $4.1$ \\
$600$ & $3.988 \times 10^{5}$ & $256 \times 64 \times 256$ & $0$ & $30$ & $3.281$ & $900$ & $0.610$ & $4.1$ \\
$90$ & $8.916 \times 10^{3}$ & $256 \times 64 \times 128$ & $0$ & $11$ & $1.034$ & $200$ & $0.433$ &  $4.3$ \\
$95$ & $9.934 \times 10^{3}$ & $256 \times 64 \times 128$ & $0$ & $11$ & $1.036$ & $200$ & $0.446$ &  $4.3$ \\
$101$ & $1.123 \times 10^{4}$ & $256 \times 64 \times 128$ & $0$ & $11$ & $1.037$ & $200$ & $0.459$ & $4.3$ \\
$105$ & $1.214 \times 10^{4}$ & $256 \times 64 \times 128$ & $0$ & $11$ & $1.039$ & $200$ & $0.468$ &  $4.3$ \\
$110$ & $1.332 \times 10^{4}$ & $256 \times 64 \times 128$ & $0$ & $11$ & $1.041$ & $200$ & $0.477$ &  $4.3$ \\
$115$ & $1.456 \times 10^{4}$ & $256 \times 64 \times 128$ & $0$ & $11$ & $1.044$ & $200$ & $0.487$ &  $4.3$ \\
$120$ & $1.585 \times 10^{4}$ & $256 \times 64 \times 128$ & $0$ & $11$ & $1.046$ & $200$ & $0.497$ &  $4.3$ \\
$125$ & $1.712 \times 10^{4}$ & $256 \times 64 \times 128$ & $0$ & $11$ & $1.051$ & $200$ & $0.508$ &  $4.3$ \\
$131$ & $1.889 \times 10^{4}$ & $256 \times 64 \times 128$ & $0$ & $11$ & $1.058$ & $200$ & $0.519$ &  $4.3$ \\
$135$ & $2.006 \times 10^{4}$ & $256 \times 64 \times 128$ & $0$ & $11$ & $1.066$ & $200$ & $0.528$ &  $4.3$ \\
$140$ & $2.157 \times 10^{4}$ & $256 \times 64 \times 128$ & $0$ & $11$ & $1.081$ & $200$ & $0.538$ &  $4.3$ \\
$145$ & $2.314 \times 10^{4}$ & $256 \times 64 \times 128$ & $0$ & $11$ & $1.137$ & $200$ & $0.556$ &  $4.3$ \\
$151$ & $2.510 \times 10^{4}$ & $256 \times 64 \times 128$ & $0$ & $11$ & $1.208$ & $200$ & $0.576$ &  $4.3$ \\
$155$ & $2.645 \times 10^{4}$ & $256 \times 64 \times 128$ & $0$ & $11$ & $1.257$ & $200$ & $0.591$ &  $4.3$ \\
$160$ & $2.818 \times 10^{4}$ & $256 \times 64 \times 128$ & $0$ & $11$ & $1.307$ & $200$ & $0.606$ &  $4.3$ \\
$165$ & $2.997 \times 10^{4}$ & $256 \times 64 \times 128$ & $0$ & $11$ & $1.365$ & $200$ & $0.626$ &  $4.3$ \\
$172$ & $3.256 \times 10^{4}$ & $256 \times 64 \times 128$ & $0$ & $11$ & $1.431$ & $200$ & $0.651$ &  $4.3$ \\
$180$ & $3.556 \times 10^{4}$ & $256 \times 64 \times 128$ & $0$ & $11$ & $1.503$ & $200$ & $0.675$ &  $4.3$ \\
$190$ & $3.974 \times 10^{4}$ & $256 \times 64 \times 128$ & $0$ & $11$ & $1.596$ & $200$ & $0.707$ &  $4.3$ \\
$200$ & $4.403 \times 10^{4}$ & $256 \times 64 \times 128$ & $0$ & $11$ & $1.680$ & $200$ & $0.738$ &  $4.3$ \\
$210$ & $4.854 \times 10^{4}$ & $256 \times 64 \times 128$ & $0$ & $11$ & $1.758$ & $200$ & $0.768$ &  $4.3$ \\
$220$ & $5.328 \times 10^{4}$ & $256 \times 64 \times 128$ & $0$ & $11$ & $1.829$ & $200$ & $0.796$ &  $4.3$ \\
$230$ & $5.823 \times 10^{4}$ & $256 \times 64 \times 128$ & $0$ & $11$ & $1.896$ & $200$ & $0.824$ &  $4.3$ \\
$1500$ & $2.492 \times 10^{6}$ & $1536 \times 128 \times 768$ & $0$ & $30$ & $5.196$ & $500$ & $0.302$ &  $4.3$ \\
$2000$ & $4.431 \times 10^{6}$ & $1536 \times 128 \times 768$ & $0$ & $30$ & $5.783$ & $500$ & $0.364$ &  $4.3$ \\
$2500$ & $6.923 \times 10^{6}$ & $1536 \times 128 \times 768$ & $0$ & $30$ & $6.424$ & $500$ & $0.436$ &  $4.3$ \\
$3000$ & $9.969 \times 10^{6}$ & $1536 \times 128 \times 768$ & $0$ & $30$ & $7.036$ & $500$ & $0.508$ &  $4.3$ \\
$3500$ & $1.357 \times 10^{7}$ & $1536 \times 128 \times 768$ & $0$ & $30$ & $7.522$ & $500$ & $0.574$ &  $4.3$ \\
$4000$ & $1.772 \times 10^{7}$ & $1536 \times 128 \times 768$ & $0$ & $30$ & $7.487$ & $500$ & $0.652$ &  $4.3$ \\
$4500$ & $2.243 \times 10^{7}$ & $1792 \times 256 \times 1024$ & $0$ & $30$ & $7.960$ & $500$ & $0.721$ &  $4.3$ \\
$5000$ & $2.769 \times 10^{7}$ & $1792 \times 256 \times 1024$ & $0$ & $30$ & $8.262$ & $500$ & $0.782$ &  $4.3$ \\
$5500$ & $3.351 \times 10^{7}$ & $1792 \times 256 \times 1024$ & $0$ & $30$ & $8.446$ & $500$ & $0.841$ &  $4.3$ \\
$6000$ & $3.988 \times 10^{7}$ & $1792 \times 256 \times 1024$ & $0$ & $30$ & $8.935$ & $500$ & $0.896$ &  $4.3$ \\
$6500$ & $4.680 \times 10^{7}$ & $1792 \times 256 \times 1024$ & $0$ & $30$ & $9.205$ & $500$ & $0.964$ &  $4.3$ \\
$7000$ & $5.427 \times 10^{7}$ & $1792 \times 256 \times 1024$ & $0$ & $30$ & $9.845$ & $500$ & $1.026$ &  $4.3$ \\
$7500$ & $6.230 \times 10^{7}$ & $1792 \times 256 \times 1024$ & $0$ & $30$ & $10.233$ & $500$ & $1.087$ &  $4.3$ \\

$400$ & $1.772 \times 10^{5}$ & $256 \times 64 \times 256$ & $18$ & $30$ & $2.099$ & $650$ & $0.467$ & $4.3$ \\
$400$ & $1.772 \times 10^{5}$ & $256 \times 64 \times 256$ & $20$ & $30$ & $2.148$ & $650$ & $0.473$ &  $4.3$ \\
$400$ & $1.772 \times 10^{5}$ & $256 \times 64 \times 256$ & $22$ & $30$ & $2.274$ & $650$ & $0.481$ &  $4.3$ \\
$400$ & $1.772 \times 10^{5}$ & $256 \times 64 \times 256$ & $24$ & $30$ & $2.307$ & $650$ & $0.487$ &  $4.3$ \\
$400$ & $1.772 \times 10^{5}$ & $256 \times 64 \times 256$ & $26$ & $30$ & $2.331$ & $650$ & $0.490$ &  $4.3$ \\
$400$ & $1.772 \times 10^{5}$ & $256 \times 64 \times 256$ & $28$ & $30$ & $2.392$ & $650$ & $0.496$ & $4.3$ \\

$500$ & $2.769 \times 10^{5}$ & $256 \times 64 \times 256$ & $18$ & $30$ & $2.367$ & $820$ & $0.547$ &  $4.3$ \\
$500$ & $2.769 \times 10^{5}$ & $256 \times 64 \times 256$ & $20$ & $30$ & $2.422$ & $820$ & $0.557$ &  $4.3$ \\
$500$ & $2.769 \times 10^{5}$ & $256 \times 64 \times 256$ & $22$ & $30$ & $2.567$ & $820$ & $0.568$ &  $4.3$ \\
$500$ & $2.769 \times 10^{5}$ & $256 \times 64 \times 256$ & $24$ & $30$ & $2.611$ & $820$ & $0.576$ &  $4.3$ \\
$500$ & $2.769 \times 10^{5}$ & $256 \times 64 \times 256$ & $26$ & $30$ & $2.648$ & $820$ & $0.582$ &  $4.3$ \\
$500$ & $2.769 \times 10^{5}$ & $256 \times 64 \times 256$ & $28$ & $30$ & $2.726$ & $820$ & $0.595$ &  $4.3$ \\

$600$ & $3.988 \times 10^{5}$ & $256 \times 64 \times 256$ & $18$ & $30$ & $2.617$ & $920$ & $0.631$ &  $4.3$ \\
$600$ & $3.988 \times 10^{5}$ & $256 \times 64 \times 256$ & $20$ & $30$ & $2.690$ & $920$ & $0.646$ &  $4.3$ \\
$600$ & $3.988 \times 10^{5}$ & $256 \times 64 \times 256$ & $22$ & $30$ & $2.839$ & $920$ & $0.657$ &  $4.3$ \\
$600$ & $3.988 \times 10^{5}$ & $256 \times 64 \times 256$ & $24$ & $30$ & $2.899$ & $920$ & $0.668$ &  $4.3$ \\
$600$ & $3.988 \times 10^{5}$ & $256 \times 64 \times 256$ & $26$ & $30$ & $2.946$ & $920$ & $0.677$ &  $4.3$ \\
$600$ & $3.988 \times 10^{5}$ & $256 \times 64 \times 256$ & $28$ & $30$ & $3.029$ & $920$ & $0.693$ &  $4.3$ \\

$700$ & $5.427 \times 10^{5}$ & $256 \times 64 \times 256$ & $18$ & $30$ & $2.837$ & $1050$ & $0.713$ &  $4.3$ \\
$700$ & $5.427 \times 10^{5}$ & $256 \times 64 \times 256$ & $20$ & $30$ & $2.932$ & $1050$ & $0.728$ &  $4.3$ \\
$700$ & $5.427 \times 10^{5}$ & $256 \times 64 \times 256$ & $22$ & $30$ & $3.084$ & $1050$ & $0.748$ &  $4.3$ \\
$700$ & $5.427 \times 10^{5}$ & $256 \times 64 \times 256$ & $24$ & $30$ & $3.170$ & $1050$ & $0.760$ &  $4.3$ \\
$700$ & $5.427 \times 10^{5}$ & $256 \times 64 \times 256$ & $26$ & $30$ & $3.218$ & $1050$ & $0.771$ &  $4.3$ \\
$700$ & $5.427 \times 10^{5}$ & $256 \times 64 \times 256$ & $28$ & $30$ & $3.312$ & $1050$ & $0.784$ &  $4.3$ \\

$800$ & $7.089 \times 10^{5}$ & $256 \times 64 \times 256$ & $18$ & $30$ & $3.045$ & $1200$ & $0.796$ &  $4.3$ \\
$800$ & $7.089 \times 10^{5}$ & $256 \times 64 \times 256$ & $20$ & $30$ & $3.139$ & $1200$ & $0.808$ &  $4.3$ \\
$800$ & $7.089 \times 10^{5}$ & $256 \times 64 \times 256$ & $22$ & $30$ & $3.307$ & $1200$ & $0.837$ &  $4.3$ \\
$800$ & $7.089 \times 10^{5}$ & $256 \times 64 \times 256$ & $24$ & $30$ & $3.411$ & $1200$ & $0.848$ &  $4.3$ \\
$800$ & $7.089 \times 10^{5}$ & $256 \times 64 \times 256$ & $26$ & $30$ & $3.477$ & $1200$ & $0.861$ &  $4.3$ \\
$800$ & $7.089 \times 10^{5}$ & $256 \times 64 \times 256$ & $28$ & $30$ & $3.533$ & $1200$ & $0.868$ &  $4.3$ \\

$900$ & $8.972 \times 10^{5}$ & $256 \times 64 \times 256$ & $18$ & $30$ & $3.240$ & $1300$ & $0.878$ &  $4.3$ \\
$900$ & $8.972 \times 10^{5}$ & $256 \times 64 \times 256$ & $20$ & $30$ & $3.340$ & $1300$ & $0.891$ &  $4.3$ \\
$900$ & $8.972 \times 10^{5}$ & $256 \times 64 \times 256$ & $22$ & $30$ & $3.473$ & $1300$ & $0.905$ &  $4.3$ \\
$900$ & $8.972 \times 10^{5}$ & $256 \times 64 \times 256$ & $24$ & $30$ & $3.445$ & $1300$ & $0.908$ &  $4.3$ \\
$900$ & $8.972 \times 10^{5}$ & $256 \times 64 \times 256$ & $26$ & $30$ & $3.458$ & $1300$ & $0.913$ &  $4.3$ \\
$900$ & $8.972 \times 10^{5}$ & $256 \times 64 \times 256$ & $28$ & $30$ & $3.447$ & $1300$ & $0.911$ &  $4.3$ \\

$1000$ & $1.108 \times 10^{6}$ & $768 \times 256 \times 384$ & $18$ & $30$ & $3.426$ & $860$ & $0.957$ &  $4.3$ \\
$1000$ & $1.108 \times 10^{6}$ & $768 \times 256 \times 384$ & $20$ & $30$ & $3.578$ & $860$ & $0.973$ &  $4.3$ \\
$1000$ & $1.108 \times 10^{6}$ & $768 \times 256 \times 384$ & $22$ & $30$ & $3.644$ & $860$ & $0.991$ &  $4.3$ \\
$1000$ & $1.108 \times 10^{6}$ & $768 \times 256 \times 384$ & $24$ & $30$ & $3.519$ & $860$ & $0.974$ &  $4.3$ \\
$1000$ & $1.108 \times 10^{6}$ & $768 \times 256 \times 384$ & $26$ & $30$ & $3.648$ & $860$ & $0.990$ &  $4.3$ \\
$1000$ & $1.108 \times 10^{6}$ & $768 \times 256 \times 384$ & $28$ & $30$ & $3.573$ & $860$ & $0.983$ &  $4.3$ \\

$1250$ & $1.731 \times 10^{6}$ & $768 \times 256 \times 384$ & $18$ & $30$ & $3.769$ & $1020$ & $1.142$ &  $4.3$ \\
$1250$ & $1.731 \times 10^{6}$ & $768 \times 256 \times 384$ & $20$ & $30$ & $3.875$ & $1020$ & $1.178$ &  $4.3$ \\
$1250$ & $1.731 \times 10^{6}$ & $768 \times 256 \times 384$ & $22$ & $30$ & $3.887$ & $1020$ & $1.176$ &  $4.3$ \\
$1250$ & $1.731 \times 10^{6}$ & $768 \times 256 \times 384$ & $24$ & $30$ & $3.875$ & $1020$ & $1.184$ &  $4.3$ \\
$1250$ & $1.731 \times 10^{6}$ & $768 \times 256 \times 384$ & $26$ & $30$ & $3.891$ & $1020$ & $1.190$ &  $4.3$ \\
$1250$ & $1.731 \times 10^{6}$ & $768 \times 256 \times 384$ & $28$ & $30$ & $3.902$ & $1020$ & $1.185$ &  $4.3$ \\

$1500$ & $2.492 \times 10^{6}$ & $768 \times 256 \times 384$ & $18$ & $30$ & $4.100$ & $830$ & $1.333$ &  $4.3$ \\
$1500$ & $2.492 \times 10^{6}$ & $768 \times 256 \times 384$ & $20$ & $30$ & $4.105$ & $830$ & $1.341$ &  $4.3$ \\
$1500$ & $2.492 \times 10^{6}$ & $768 \times 256 \times 384$ & $22$ & $30$ & $4.108$ & $830$ & $1.353$ &  $4.3$ \\
$1500$ & $2.492 \times 10^{6}$ & $768 \times 256 \times 384$ & $24$ & $30$ & $4.228$ & $830$ & $1.356$ &  $4.3$ \\
$1500$ & $2.492 \times 10^{6}$ & $768 \times 256 \times 384$ & $26$ & $30$ & $4.214$ & $830$ & $1.364$ &  $4.3$ \\
$1500$ & $2.492 \times 10^{6}$ & $768 \times 256 \times 384$ & $28$ & $30$ & $4.120$ & $830$ & $1.348$ &  $4.3$ \\

$2000$ & $4.431 \times 10^{6}$ & $1536 \times 128 \times 768$ & $18$ & $30$ & $4.750$ & $400$ & $0.350$ &  $4.3$ \\
$2000$ & $4.431 \times 10^{6}$ & $1536 \times 128 \times 768$ & $20$ & $30$ & $4.977$ & $400$ & $0.356$ &  $4.3$ \\
$2000$ & $4.431 \times 10^{6}$ & $1536 \times 128 \times 768$ & $22$ & $30$ & $5.076$ & $400$ & $0.359$ &  $4.3$ \\
$2000$ & $4.431 \times 10^{6}$ & $1536 \times 128 \times 768$ & $24$ & $30$ & $5.199$ & $400$ & $0.359$ &  $4.3$ \\
$2000$ & $4.431 \times 10^{6}$ & $1536 \times 128 \times 768$ & $26$ & $30$ & $5.199$ & $400$ & $0.362$ &  $4.3$ \\
$2000$ & $4.431 \times 10^{6}$ & $1536 \times 128 \times 768$ & $28$ & $30$ & $5.438$ & $400$ & $0.382$ &  $4.3$ \\

$3000$ & $9.969 \times 10^{6}$ & $1536 \times 128 \times 768$ & $18$ & $30$ & $5.718$ & $300$ & $0.483$ &  $4.3$ \\
$3000$ & $9.969 \times 10^{6}$ & $1536 \times 128 \times 768$ & $20$ & $30$ & $5.934$ & $300$ & $0.490$ &  $4.3$ \\
$3000$ & $9.969 \times 10^{6}$ & $1536 \times 128 \times 768$ & $22$ & $30$ & $6.142$ & $300$ & $0.496$ &  $4.3$ \\
$3000$ & $9.969 \times 10^{6}$ & $1536 \times 128 \times 768$ & $24$ & $30$ & $6.331$ & $300$ & $0.509$ &  $4.3$ \\
$3000$ & $9.969 \times 10^{6}$ & $1536 \times 128 \times 768$ & $26$ & $30$ & $6.529$ & $300$ & $0.511$ &  $4.3$ \\
$3000$ & $9.969 \times 10^{6}$ & $1536 \times 128 \times 768$ & $28$ & $30$ & $6.525$ & $300$ & $0.518$ &  $4.3$ \\

$400.9 \times 10^{3}$ & $2.440 \times 10^{5}$ & $100 \times 50 \times 50$ & $0$ & $2\pi$ & $2.942$ & $1688$ & $3.191$ &  $\text{A}.3$ \\
$400.9 \times 10^{3}$ & $2.440 \times 10^{5}$ & $200 \times 100 \times 100$ & $0$ & $2\pi$ & $2.911$ & $252$ & $1.557$ &  $\text{A}.3$ \\
$400.9 \times 10^{3}$ & $2.440 \times 10^{5}$ & $300 \times 150 \times 150$ & $0$ & $2\pi$ & $2.994$ & $744$ & $1.089$ &  $\text{A}.3$ \\
$400.9 \times 10^{3}$ & $2.440 \times 10^{5}$ & $300 \times 150 \times 300$ & $0$ & $2\pi$ & $2.995$ & $790$ & $0.447$ &  $\text{A}.3$ \\
$400.9 \times 10^{3}$ & $2.440 \times 10^{5}$ & $400 \times 300 \times 400$ & $0$ & $2\pi$ & $2.995$ & $306$ & $0.438$ &  $\text{A}.3$ \\

\hline

\end{longtable}

\bibliographystyle{jfm}
\bibliography{bibliothek}

\end{document}